\documentclass[prb,twocolumn,unsortedaddress,floatfix]{revtex4}

\usepackage{dcolumn,amsmath,xspace}
\usepackage{graphicx}
\usepackage{dcolumn}
\usepackage{amssymb}
 \usepackage{latexsym}

\usepackage{units}

\newcommand{\sixrt}{\ensuremath{(6\sqrt{3}\!\times\!6\sqrt{3})\text{R}30^\circ}~}

\begin{document}
\title{Structure and electronic properties of epitaxial graphene grown on SiC\\}

\author{M. Sprinkle}
\author{J. Hicks}
\affiliation{The Georgia Institute of Technology, Atlanta, Georgia 30332-0430, USA}
\author{A. Tejeda}
\affiliation{Institut Jean Lamour, CNRS - Univ. de Nancy - UPV-Metz, 54506 Vandoeuvre les Nancy, France}
\affiliation{Synchrotron SOLEIL, L'Orme des Merisiers, Saint-Aubin, 91192 Gif sur Yvette, France}
\author{A. Taleb-Ibrahimi}
\affiliation{UR1 CNRS/Synchrotron SOLEIL, Saint-Aubin, 91192 Gif sur Yvette, France}
\author{P. Le F\`{e}vre}
\author{F. Bertran}
\affiliation{Synchrotron SOLEIL, L'Orme des Merisiers, Saint-Aubin, 91192 Gif sur Yvette, France}
\author{H. Tinkey}
\author{M.C. Clark}
\affiliation{The Georgia Institute of Technology, Atlanta, Georgia 30332-0430, USA}
\author{P. Soukiassian}
\affiliation{Commissariat\`{a} l'Energie Atomique, SIMA, DSM-IRAMIS-SPCSI, Saclay, 91191 Gif sur Yvette, France}
\affiliation{D\'{e}pt. de Physique, Univ. de Paris-Sud, 91405 Orsay, France}
\author{D. Martinotti }
\affiliation{Commissariat\`{a} l'Energie Atomique, SIMA, DSM-IRAMIS-SPCSI, Saclay, 91191 Gif sur Yvette, France}
\author{J. Hass}
\author{W.A. de Heer}
\affiliation{The Georgia Institute of Technology, Atlanta, Georgia 30332-0430, USA}
\author{C. Berger}
\affiliation{The Georgia Institute of Technology, Atlanta, Georgia 30332-0430, USA}
\affiliation{CNRS/Institut N\'{e}el, BP166, 38042 Grenoble, France}
\author{E.H. Conrad}
\affiliation{The Georgia Institute of Technology, Atlanta, Georgia 30332-0430, USA}

\begin{abstract}
We review progress in developing epitaxial graphene as a material for carbon electronics.  In particular, improvements in epitaxial graphene growth, interface control and the understanding of multilayer epitaxial graphene's electronic properties are discussed. Although graphene grown on both polar faces of SiC is addressed, our discussions will focus on graphene grown on the $(000\bar{1})$ C-face of SiC. The unique properties of C-face multilayer epitaxial graphene have become apparent.  These films behave electronically like a stack of nearly independent graphene sheets rather than a thin Bernal-stacked graphite sample. The origin of multilayer graphene's electronic behavior is its unique highly-ordered stacking of non-Bernal rotated graphene planes.  While these rotations do not significantly affect the inter-layer interactions, they do break the stacking symmetry of graphite.  It is this broken symmetry that causes each sheet to behave like an isolated graphene plane.
\end{abstract}
\vspace*{4ex}

\maketitle
\newpage

\section{Introduction\label{S:Intro}}
Although graphene was first isolated by chemical exfoliation in 1961~\cite{Boehm_ZNature_62} and later shown to grow epitaxially on SiC in 1975~\cite{vanBommel75} and 1998,~\cite{Forbeaux_PRB_98} research on graphene's electronic properties did not begin until 2001.\cite{de Heer_01,Berger04} As is so often promoted, the driving motivation for this research explosion is graphene's potential for carbon electronics. For this end game to be reachable, there are two important properties of graphene that must be achieved.  First, no matter how graphene is produced, it must have the band structure (and thus the transport properties) of an ideal isolated graphene sheet.  Second any method of graphene production and device fabrication must be scalable up from a single nanometer scale prototype to macroscopic systems in which millions of graphene switches and their interconnects are integrated. Current research has focused on three ways of producing graphene: (i) epitaxial graphene grown on SiC,\cite{Charrier_JAP_02,Berger04,Hass_JPhyCM_08,Virojanadara_PRB_08,Emtsev_NatM_09,Tromp_PRL_09} (ii) graphene grown on metals,\cite{Obraztsov_Nano_let_09} and (iii) exfoliated graphene separated from bulk graphite crystals and deposited on $\text{SiO}_2$ substrates.\cite{Novoselov_Science_04,Stankovich_Carbon_07} The latter two methods both require an end step: the transfer of graphene to a semiconducting or insulating device platform. This transfer step is recognizable as the same problem that has plagued the development of carbon nanotube (CNT) electronics for the past two decades.  In fact, the original impetus to develop epitaxial graphene electronics was to take advantage of CNT's unique electronic properties while finding a way to circumvent CNT's lack of scalability.\cite{Berger04,Berger06}  The scalability of epitaxial graphene is well recognized.  It has already been used to construct multi-FET arrays~\cite{Kedzierski_IEEE_08} and has recently been used to build a 10,000 FET transistor array on a $3.5\!\times\!4.5$mm SiC substrate.\cite{Sprinkle_TBP_sidewall10k}
\begin{figure}
\includegraphics[width=5.5cm,clip]{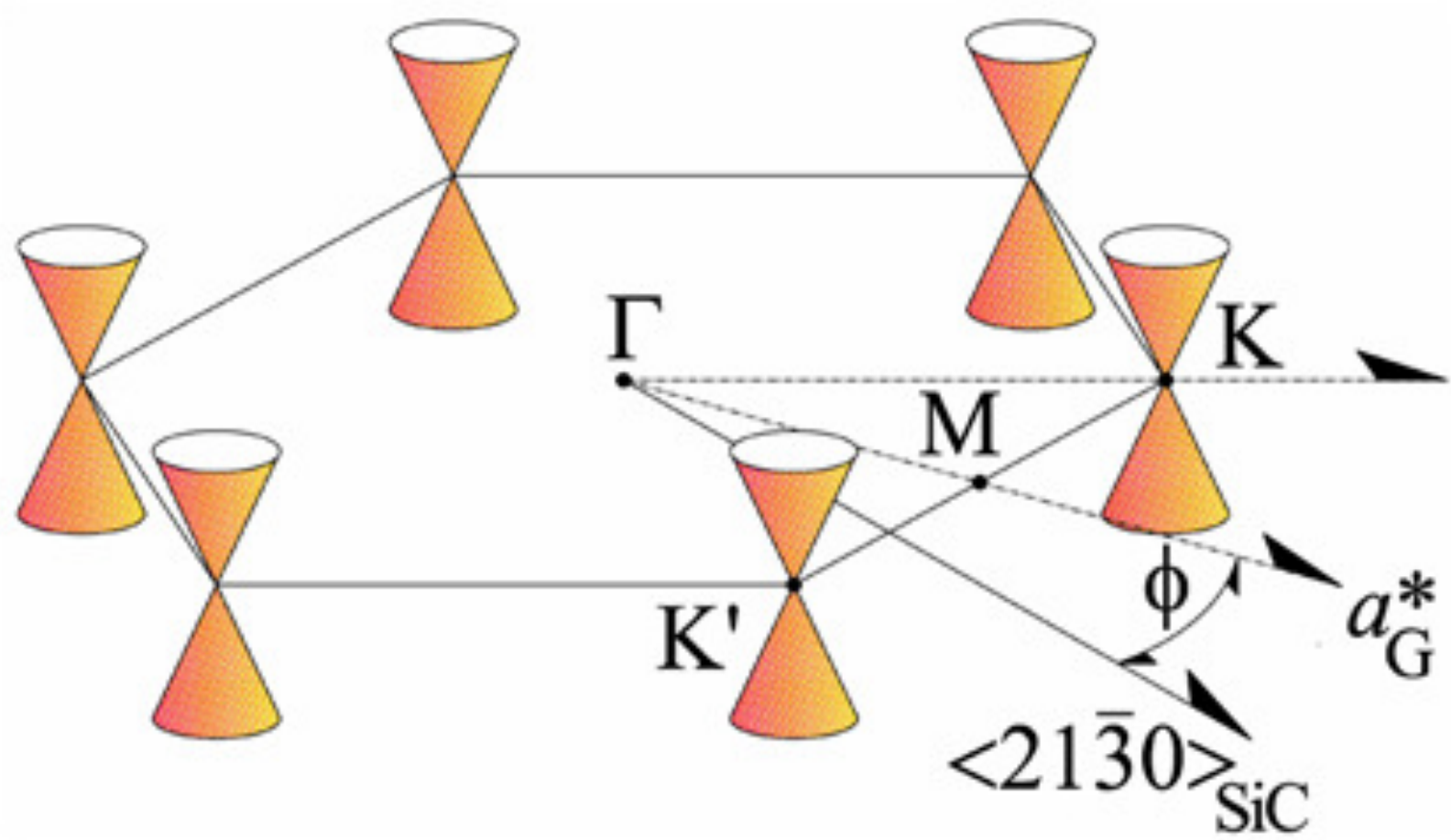}
\caption{2D Brillouin zone of graphene near $E_F$ showing the six Dirac cones at the $K$-points. The cones are shown rotated through an angle $\phi$ relative to the SiC $\langle 21\bar{3}0\rangle$ direction.} \label{F:Brillouin_Z}
\end{figure}

In addition to its scalability, epitaxial graphene behaves electronically like an isolated graphene sheet.\cite{Berger06,Sprinkle_PRL_09,Sprinkle_PSSRRL_09,Riedl_tbp_arXiv_09,de Heer_SSC_07,Sadowski06,Sadowski07a,Wu_PRL_07,Orlita_PRL_08,Miller_Science_09} Its relevant electronic signature is the dispersion of the $\pi$- and $\pi^*$-bands near the six $K$-points of the graphene hexagonal reciprocal unit cell.\cite{graphene_BS_ref} The dispersion is linear $E(\Delta k)\!=\!\hbar v_F \Delta k$; where $v_F\sim\!10^6\text{m/sec}$ is the Fermi velocity and $\Delta k$ is the momentum relative to the $K$-points. The two-dimensional dispersion is isotropic and defines a cone (referred to as a Dirac cone) with an apex at the Dirac point, $E_D$.\cite{graphene_BS_ref} For undoped graphene the Fermi energy, $E_F$, coincides with $E_D$ so that the Fermi surface consists of six points [see Fig.~\ref{F:Brillouin_Z}]. This unique electronic structure is relevant for graphene based electronics for several reasons. For instance, the magnitude of $v_F$ means that electrons with energies significantly larger than thermal energies ($\sim\!1\text{eV}$) relative to $E_D$  have wavelengths of the order of $4\pi\hbar v_F/1\text{eV}\!\sim\!2$nm. Consequently quantum confinement becomes important in nanoscopic graphene structures with these dimensions and can lead to band gaps of the order of $\sim\!1$eV.\cite{Berger04}  In addition the symmetry change of the electron wave function as an electron moves from $K$ to $K'$, is responsible for the ballistic transport properties of graphene.\cite{de Heer_SSC_07}

While most forms of graphene show many of the properties of an isolated graphene sheet, only multilayer epitaxial graphene (MEG) grown on the C-face of SiC exhibits them all.\cite{Berger06,Sadowski06,Sadowski07a,Wu_PRL_07,de Heer_SSC_07,Orlita_PRL_08,Miller_Science_09,Sprinkle_PRL_09} In addition, Landau level spectroscopy from MEG films has demonstrated unprecedented graphene properties including exceptionally high room temperature mobilities ($>\!250,000\text{cm}^2$/Vs), resolved Landau levels in magnetic fields as low as 50 mT and remarkably low electron-phonon coupling up to room temperature.\cite{Miller_Science_09}

Despite these impressive properties, it is often incorrectly stated that epitaxial graphene is not true  isolated graphene. The arguments imply that because the $\pi$- and $\pi^*$ bonds responsible for electron transport are perpendicular to the graphene plane, it is reasonable to conclude that either placing graphene on any substrate or stacking graphene in multiple sheets will have a significant impact on its band structure near the $K$-points.  While this can be true, the literature oversimplifies the actual physics of the graphene-SiC substrate or even the graphene-graphene interaction. This thinking imposes unnecessary constraints on a number of potential research avenues and can actually hinder the pace of developing graphene electronics.  Therefore, before discussing the details of epitaxial graphene, it is worth taking some time to clarify these issues.

\subsection{Graphene-substrate interactions}
Graphene tends to interact more strongly with most materials than with itself.  In fact this is the reason that graphene can be exfoliated onto a $\text{SiO}_2$ surface; the graphene adheres to the $\text{SiO}_2$ strong enough to allow one or multiple layers to cleave from bulk graphite.  The question is not if graphene bonds to a surface but how the bonding affects the electronic properties of the film.  In the case of exfoliated graphene on $\text{SiO}_2$ the graphene substrate bond has not been extensively studied.  This is due in part to the problem of finding small flakes with standard surface analysis probes.  Even in the case where the band structure of exfoliated films has been directly probed by angle resolved photoemission spectroscopy (ARPES),  the film disorder severely reduces the $k$-resolution of the technique and makes quantitative analysis problematic.\cite{Knox_PRB_08} Transport measurements indirectly show the effects of the substrate: large spatial charge fluctuations and electron-doping levels as high as 300meV.\cite{Martin_NPYS_08,Knox_PRB_08}  Indeed many of these problems can be reduced by an order of magnitude simply by removing the substrate and suspending exfoliated graphene over etched channels.\cite{Bolotin_SSC_08,Du_NatureNano_08} The change in transport and doping between graphene with and without the $\text{SiO}_2$ substrate in part demonstrates the influence of the substrate interaction.

However, not all substrate interactions are created equal. Understanding the differences in how graphene interacts with its environment, has important consequences in how to manipulate graphene's transport properties.
\begin{figure}
\includegraphics[width=7.5cm,clip]{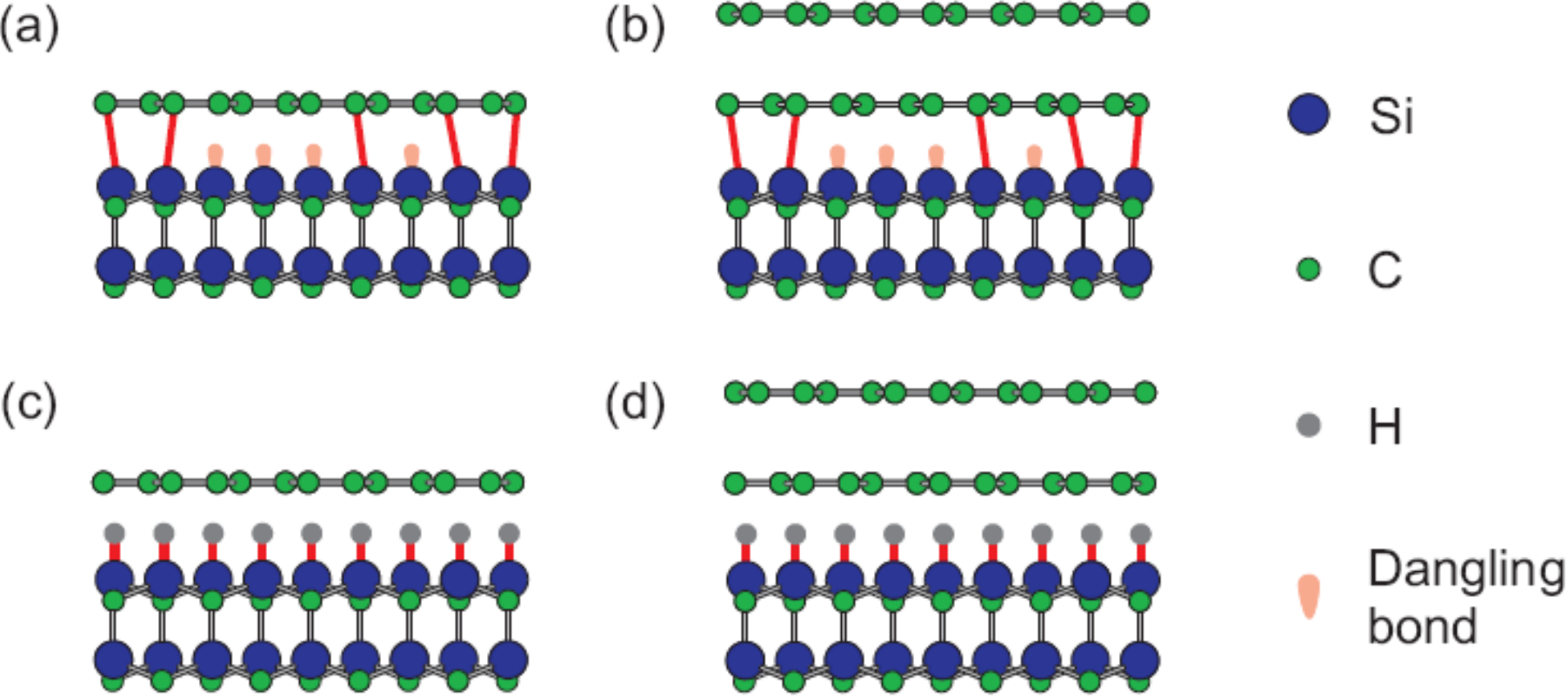}
\caption{Schematic models of graphene on the SiC Si-face. (a) Strongly bonded graphene zero-layer (ZL) or buffer layer. (b) graphene layer above the buffer. (c) $\text{H}_2$ passivated buffer layer.  (d)  $\text{H}_2$ passivated surface with two graphene layers. From Ref. [\onlinecite{Riedl_tbp_arXiv_09}].} \label{F:Buffer_model}
\end{figure}
\begin{figure}
\includegraphics[width=7.0cm,clip]{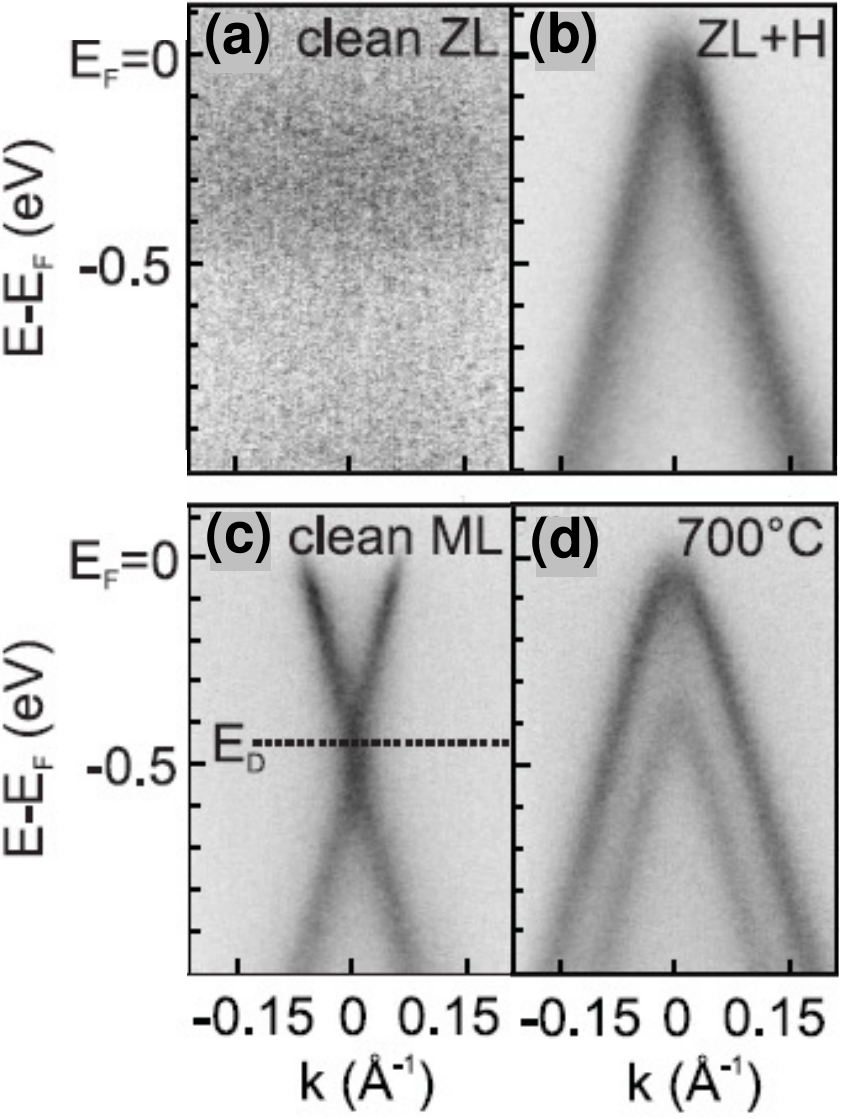}
\caption{ARPES data near the $K-$point of graphene grown on the Si-face of SiC. (a) for ZL graphene there are no observable bands. (b) Dirac cone from the ZL graphene after $\text{H}_2$ passivation. (c) Doped Dirac cone from a graphene layer above the ZL. (d) {\it AB} bilayer bands from ZL and the first graphene layer in (c) after $\text{H}_2$ passivation of the interface. From Ref.~[\onlinecite{Riedl_tbp_arXiv_09}].} \label{F:H2_passivation}
\end{figure}

For epitaxial graphene grown on either the SiC(0001) Si-terminated face (Si-face) or the SiC$(000\bar{1})$ C-terminated face (C-face), $\text{sp}^3$ bonding causes a strong graphene-substrate interaction.\cite{Hass_JPhyCM_08,Starke_JPCM_09}  In fact the interaction is so strong on the Si-face that the graphene interface layer, sometimes called the zeroth layer (ZL) or buffer layer, becomes a wide gap semiconductor with none of the electronic character of isolated graphene [see Fig.~\ref{F:Buffer_model}(a) and Fig.~\ref{F:H2_passivation}(a)].\cite{Varchon_PRL_07} Nonetheless, the existence of this buffer layer has a major advantage; it largely decouples subsequent graphene layers from the substrate.  This has been demonstrated very nicely on Si-face epitaxial graphene where the linear bands of graphene are intact (although doped) in the layer above the buffer layer [see Fig.~\ref{F:Buffer_model}(b) and Fig.~\ref{F:H2_passivation}(c)].\cite{Ohta_PRL_07}

While the buffer layer preserves the linear bands, the graphene layers are still n-doped by as much as 0.44eV on the Si-face\cite{Ohta_PRL_07} and 0.2eV on the C-face\cite{Emtsev_PRB_08} due to substrate charge transfer. It is not clear if the charging is from an intrinsic Schottky barrier, defect SiC dangling bond states at the interface or both.\cite{Varchon_PRL_07} It is clear, however, that these states can be removed by hydrogen passivation.  Riedl et al.\cite{Riedl_tbp_arXiv_09} have been able to intercalate molecular hydrogen between the Si-face SiC and the buffer layer, which simultaneously breaks the ZL graphene SiC $\text{sp}^3$ bonds and saturates the remaining SiC dangling bonds [see Figs.~\ref{F:Buffer_model}(c) and (d)]. These experiments are a beautiful example of how interface manipulation can eliminate the charge transfer to the first few graphene layers and restore the linear dispersion of the buffer layer in epitaxial graphene [see Fig.~\ref{F:H2_passivation}(b)].

\subsection{Symmetry and graphene-graphene interactions\label{S:Symmetry}}
The role of symmetry in determining the band structure of graphene sheets can be demonstrated by comparing graphene with different stacking arrangements.  In graphite the graphene sheets are rotated $60^\circ$ relative to adjacent sheets in the stack (Bernal stacking).\cite{Hass_JPhyCM_08}   Bernal stacking causes the two atoms per graphene cell to be inequivalent (the two atoms are labeled `{\it A}' and `{\it B}').  The `{\it A}' atoms are bonded to an atom in the graphene sheets above and below through $p_z$ orbitals.   The `{\it B}'  atoms lie in the center of graphene hexagons in sheets above and below and are therefore not bonded to other carbon atoms.  The stacking induced in-equivalence of the `{\it A}'  and `{\it B}' atoms in graphite is often referred to as {\it AB} stacking.  

Si-face graphene is Bernal stacked like graphite.\cite{Ohta_PRL_07} This has been elegantly demonstrated by comparing Si-face graphene that has been grown with both a ZL and an additional graphene layer before and after $\text{H}_2$ passivation.\cite{Riedl_tbp_arXiv_09} In the as-grown Si-face graphene, only the top layer is isolated and exhibits the band structure of a single graphene sheet [see Figs.~\ref{F:H2_passivation}(c)]. When the $\text{H}_2$ is intercalated into the interface, the ZL becomes isolated from the substrate and the two graphene layers become a new electronic system.  Because the ZL is rotated $60^\circ$ relative to the top layer (Bernal stacking), the doped Dirac cone of the single layer becomes the split bilayer bands structure of a bilayer pair [see Fig.~\ref{F:H2_passivation}(d)].

While it is tempting to assume that the band structure change is due to the weak bonding of the $\pi$ orbitals between planes, this is incorrect. The change is instead due to the symmetry change of the bilayer system.  To see this we can examine the band structure of bi-layer graphene when interplanar interactions exist but the {\it AB} stacking symmetry is destroyed. The simplest example is {\it AA} stacking.\cite{Hass_JPhyCM_08}  In {\it AA} stacking ($0^\circ$ rotation) the number of bonds/area is double that of Bernal stacking (one for every atom). Despite this increase, the dispersion in {\it AA} stacking reverts back to the linear bands of graphene.\cite{Charlier_PRB_92} This effect is more general and applies to any rotation other than $60^\circ$.

\begin{figure}
\includegraphics[angle=90, width=7.5cm,clip]{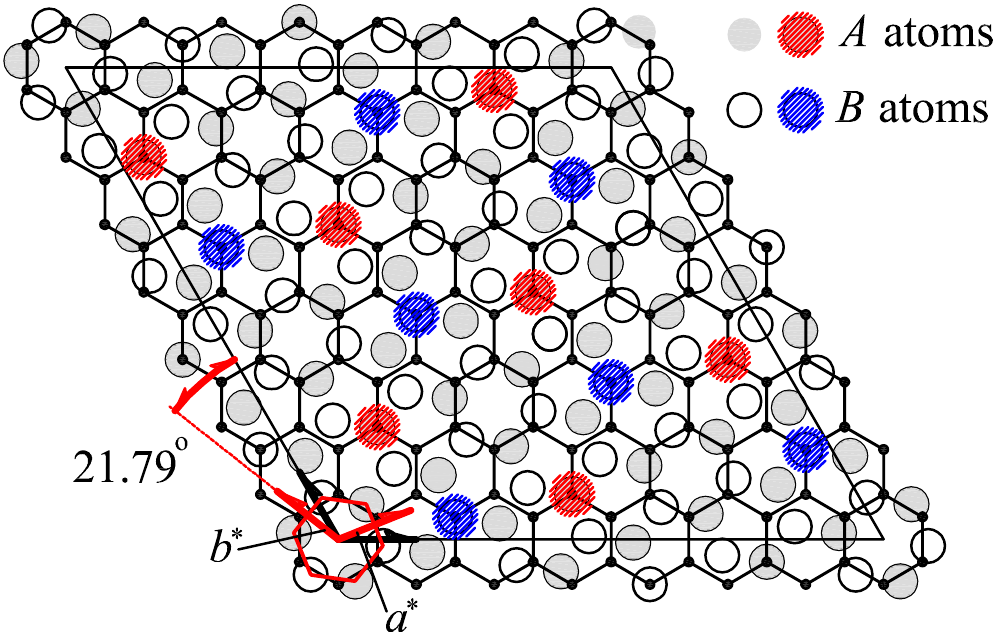}
\caption{A model of a $7\!\times\! 7$ commensurate structure produced by two graphene sheets with a relative rotation of $21.79^\circ$.  The top sheet is represented by the ball and stick model. The lower sheet is represented by open and grey circles.  In the lower layer {\it A} atoms (open circles) and {\it B} atoms (grey circles) are directly below and bonded to atoms in the upper layer (shaded red or blue) with equal probability.  Although not easily visable, shaded and open atoms in the layer below occupy positions in the upper lattice with equal probability.} \label{F:Morie_model}
\end{figure}

\begin{figure}
\begin{center}
\includegraphics[width=7.0cm,clip]{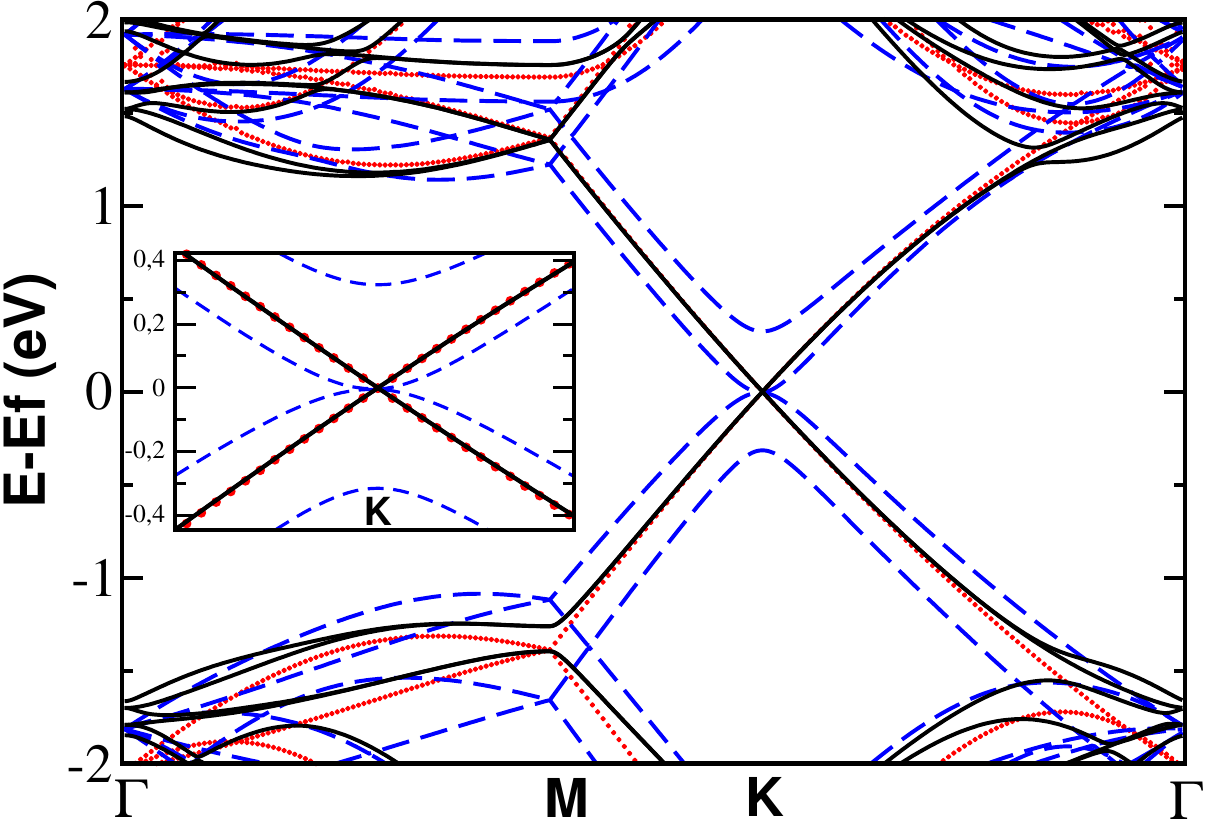}
\caption{Calculated band structure for three forms of graphene. (i) isolated graphene sheet (dots), (ii) {\it AB} graphene bi-layer (dashed line) and (iii) a $(\sqrt{13}\times\sqrt{13})_\text{G}\text{R}46.10^\circ$ structure formed by two stacked graphene sheets with a relative rotation angle of $32.204^\circ$ (solid line). Note that subscript ``G'' refers to graphene lattice vectors. Inset shows the details of band structure at the $K$-point. From Ref.~[\onlinecite{Hass_PRL_08}].} \label{F:Band_structure}
\end{center}
\end{figure}

Although non-$60^\circ$ rotations lead to small increases in the bonding energy per atom (of the order of a few meV/atom),\cite{Kolmogorov_PRB_05,Campanera_PRB_07} these small energetics concerns are overshadowed by a more important symmetry change that leads to dramatic changes in the electronic band structure of rotated graphene sheets.  Here rotation means that the relative angle between adjacent sheets is some value other than $60^\circ$.  To demonstrated this, we show a ball model of two graphene sheets rotated by $21.79^\circ$ in Fig.~\ref{F:Morie_model} (a similar structure can also be formed by rotating $16.43^\circ$).  For the commensurate structure shown in Fig.~\ref{F:Morie_model}, the in-equivalency of the `{\it A}' and `{\it B}' atoms, associated with Bernal stacking, is not longer true.  There are as many `{\it A}' atoms in sites below atoms in the upper plane (bonding sites) as there are `{\it B}'  atoms in the same position.  Actually, all `{\it A}' atoms in the layer below occupy positions in the upper graphene lattice equally likely as `{\it B}' atoms. In other words there is no symmetry breaking in the sheet so `{\it A}' and `{\it B}' atoms are essentially equivalent in terms of bonding and structure.  This result is true for any rotation angle other than $60^\circ$.  The effect of this symmetry is dramatic.  Both planes become electronically equivalent to an isolated graphene sheet. This has been demonstrated theoretically a number of ways for both small and large relative rotation angles.\cite{LopesdoSantos_PRL_07,Latil_PRB_07,Hass_PRL_08}  Figure \ref{F:Band_structure} compares the calculated band structure for two graphene sheets rotated by $32.204^\circ$.  The electronic bands are obviously different from the {\it AB} stacked pair and indistinguishable near the Dirac point from the bands of a single graphene layer.

In C-face graphene the majority stacking is non-graphitic so that films as thick as 60 graphene layers still behave electronically like a stack of isolated graphene sheets.\cite{Sadowski07a,Orlita_PRL_08}  In Sec.~\ref{S:Struc} we will show that {\it AB} stacking is the exception rather than the rule in C-face films and that the band structure of these multilayer films are equivalent to isolated graphene. 

\section{Structure and Stacking\label{S:Struc}}
Graphene grown on the C-face of SiC is significantly different than Si-face graphene, both in how it grows and its structural order.\cite{Hass_JPhyCM_08} While the slow growth rates of Si-face graphene makes it relatively easily to grow 1-3 layers films, the high growth rates in C-face graphene makes thin film growth much more difficult. In addition, Si-face epitaxial graphene grows Bernal stacked (i.e, with a $60^\circ$ relative rotation between adjacent planes), while C-face graphene grows in an ordered set of relative rotational angles. In this section we will focus on the detailed morphology and electronic band structure of C-face multilayer epitaxial graphene films.  In Sec. \ref{S:Stacking} we will show that the rotational stacking in C-face graphene is highly ordered and is not disordered as is often asserted in the literature.

\subsection{Structure of C-face graphene\label{S:Struc_C-face}}
\begin{figure}
\includegraphics[width=8.0cm,clip]{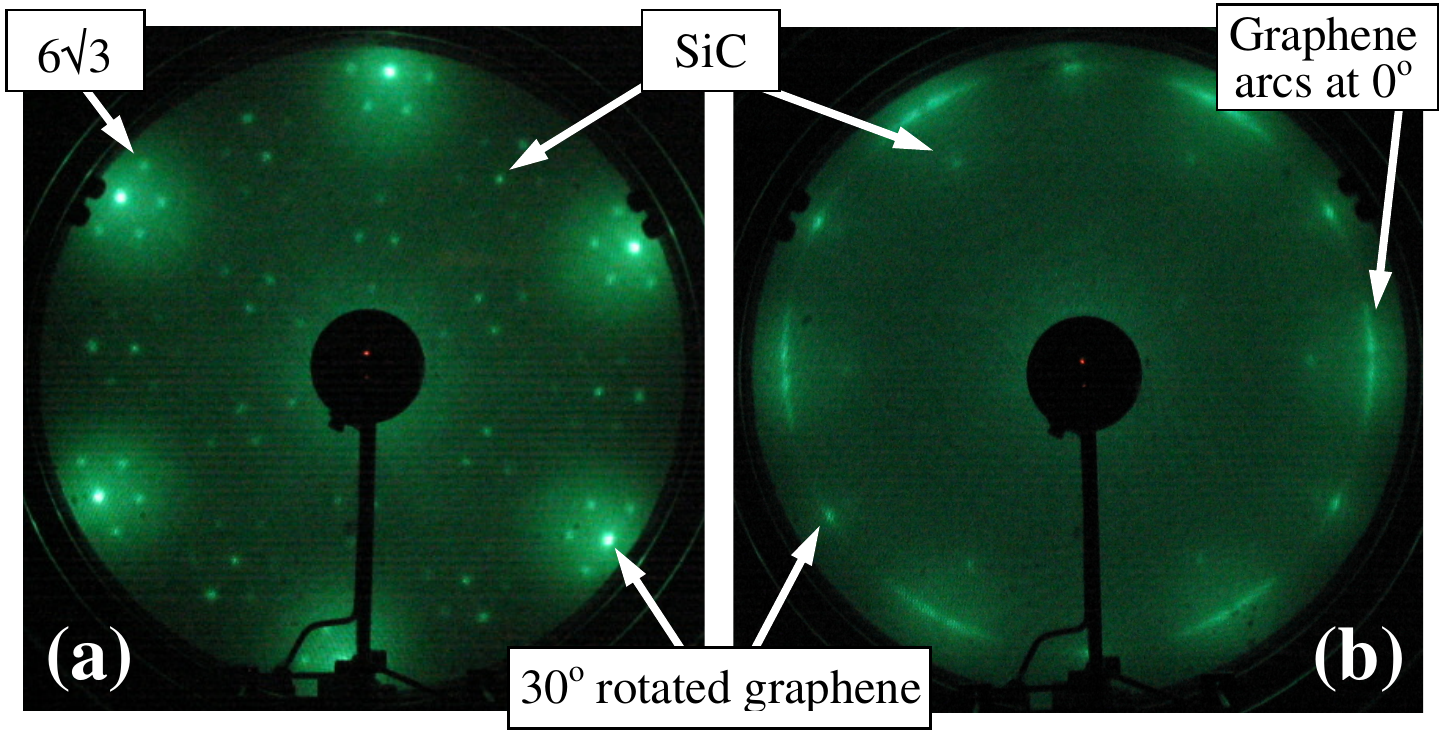}
\caption{LEED patterns from (a) Si- and (b) C-face graphene films grown on SiC.  Both images show SiC diffraction spots.  In (a) the graphene is rotated $30^\circ$ relative to SiC and the \sixrt reconstruction spots are visible. In (b) a set of six diffuse arcs rotated $0^\circ$ from SiC are also visible.} \label{F:LEED_Si_C}
\end{figure}

\begin{figure}
\includegraphics[width=6.5cm,clip]{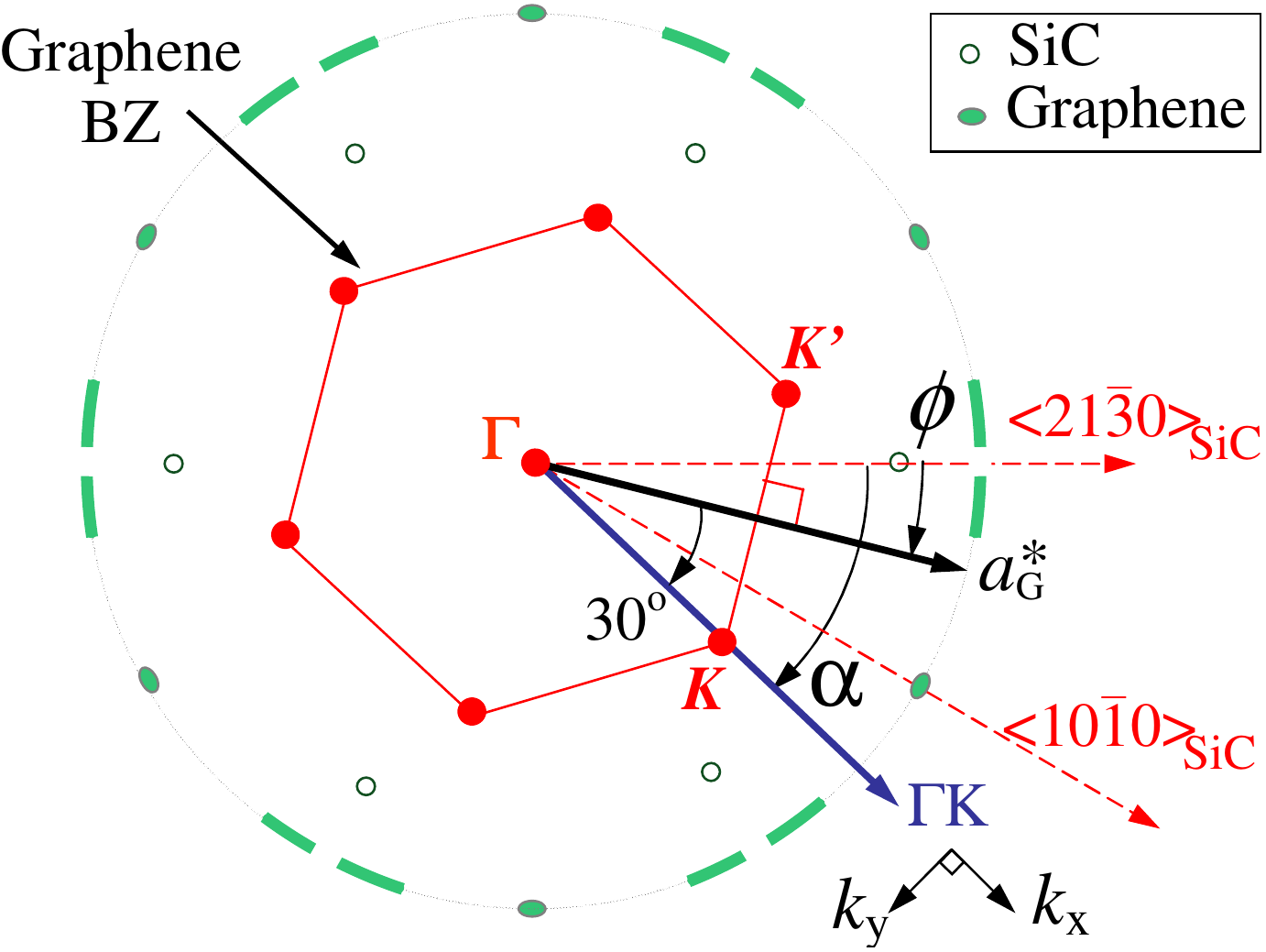}
\caption{Schematic LEED pattern from a C-face graphene film.  Both graphene spots rotated $30^\circ$ from the SiC $\langle21\bar{3}0\rangle$ direction and diffuse graphene arcs centered around $\phi\!=0^\circ$ are shown.  A hexagonal graphene Brillouin Zone (BZ) rotated by $\phi$ is shown. Note that the $\Gamma K$ direction of the graphene BZ is rotated $30^\circ$ relative to the graphene reciprocal lattice vector $a^*_G$.} \label{F:Schem_LEED}
\end{figure}

Graphene grown on the Si-face is always oriented $30^\circ$ relative to the SiC $\langle21\bar{3}0\rangle$ direction and leads to a \sixrt reconstruction in low energy electron diffraction (LEED) images [see Fig.~\ref{F:LEED_Si_C}(a)].\cite{Hass_JPhyCM_08}  Graphene grown on the C-face is also rotated $30^\circ$ relative to the $\langle21\bar{3}0\rangle$ direction but LEED and surface x-ray diffraction (SXRD) indicates that some planes are rotated within a small angular region around the SiC $\langle21\bar{3}0\rangle$ [see Fig.~\ref{F:LEED_Si_C}(b) and Fig.~\ref{F:Schem_LEED}].\cite{Hass_PRL_08,Hass_JPhyCM_08}

\begin{figure}
\includegraphics[width=8.0cm,clip]{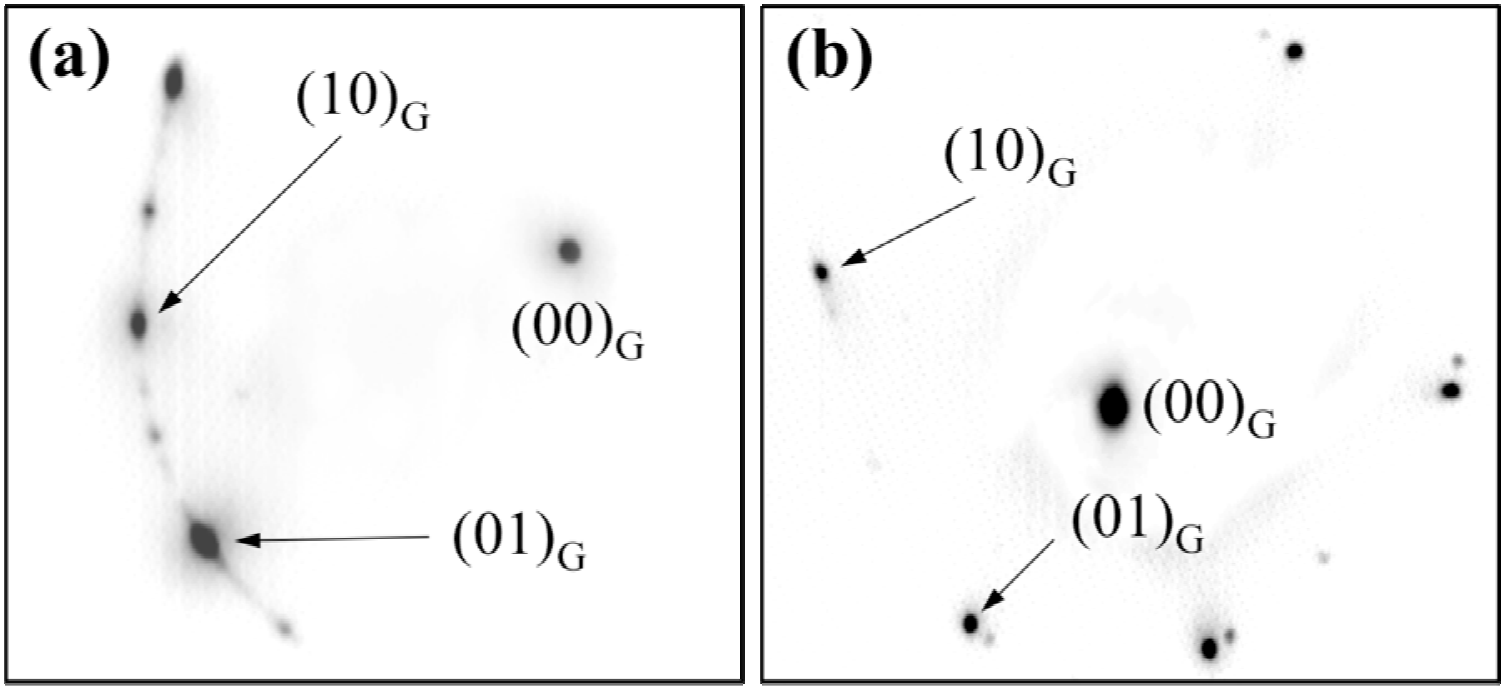}
\caption{LEEM micro LEED images of (a) a UHV grown 3- to 4-layer C-face graphene film.  Only one quadrant of the pattern is shown but the two primary graphene reflections are given for reference.  Note diffuse arcs and extra graphene spots.  (b)  a furnace grown 11-layer graphene film on  the C-face of SiC.  Only two rotated graphene planes are visible.  The image is slightly distorted from aberrations in the LEEM.} \label{F:UHV_vs_Furn}
\end{figure}

The additional rotation angles are seen in both Ultra High Vacuum (UHV) and furnace grown graphene. However, it is important to recognize that there are significant morphological differences in the graphene sheets grown in UHV and at ambiant pressure furnace.  In Fig.~\ref{F:UHV_vs_Furn} we compare low energy electron microscopy (LEEM) micro LEED images of UHV and furnace grown graphene. The image for a UHV film shows both extra rotated graphene spots and diffuse graphene arcs.  This is evidence of a large number of rotation angles within the $\sim 1\mu$m beam diameter, consistent with STM studies on UHV samples.\cite{Varchon_PRB_77_08}  Furnace grown samples, on the other hand, show only a few different graphene rotations in the same area.  This emphasizes that while both methods produce rotated sheets, the size of the sheets in UHV films is very small.\cite{Varchon_PRB_77_08,Biedermann_PRB_09}  In the furnace grown samples, the size of the graphene sheets is much larger.  In fact, grain boundaries have yet to be seen in furnace grown graphene, suggesting that continuous graphene sheets can span macroscopic dimensions.

It is very important to realize that the C-face rotated diffraction patterns, like the one shown in Fig.~\ref{F:LEED_Si_C}(b), are not due to randomly rotated domains of graphite crystals like those in HOPG graphite.  Instead, the rotated sheet are interleaved in the multilayer graphene stack.\cite{Hass_PRL_08,Sprinkle_PRL_09,Sprinkle_PSSRRL_09}  This statement is the result of a wide variety of experiments including transport measurements, electronic structure and structural studies. These will be discussed in detail below and in Sec.~\ref{S:Stacking}.  Because of C-face graphene's unique electronic properties and scalability, we will focus on furnace grown C-face graphene in the rest of this review.

To begin to understand the rotational structure and its implication to C-face graphene's electronic properties, we begin by correlating SXRD and ARPES data.  To make this easier Fig.~\ref{F:Schem_LEED} shows a schematic LEED pattern from a C-face graphene film like the LEED pattern in Fig.~\ref{F:LEED_Si_C}(b).  For reference a graphene sheet (with a reciprocal lattice vector $a_G^*$) rotated by an angle $\phi$ relative to the SiC $\langle21\bar{3}0\rangle$ produces diffraction intensity on the graphene arc at the same angle $\phi$.  The corresponding graphene Brillouin zone (BZ) is rotated by $\phi$ so that the $\Gamma K$ direction vector is rotated by an angle $\alpha\!=\!\phi\!+\!30^\circ$ relative to the SiC $\langle21\bar{3}0\rangle$ direction [see Fig.~\ref{F:Schem_LEED}].

\begin{figure}
\includegraphics[width=7.5cm,clip]{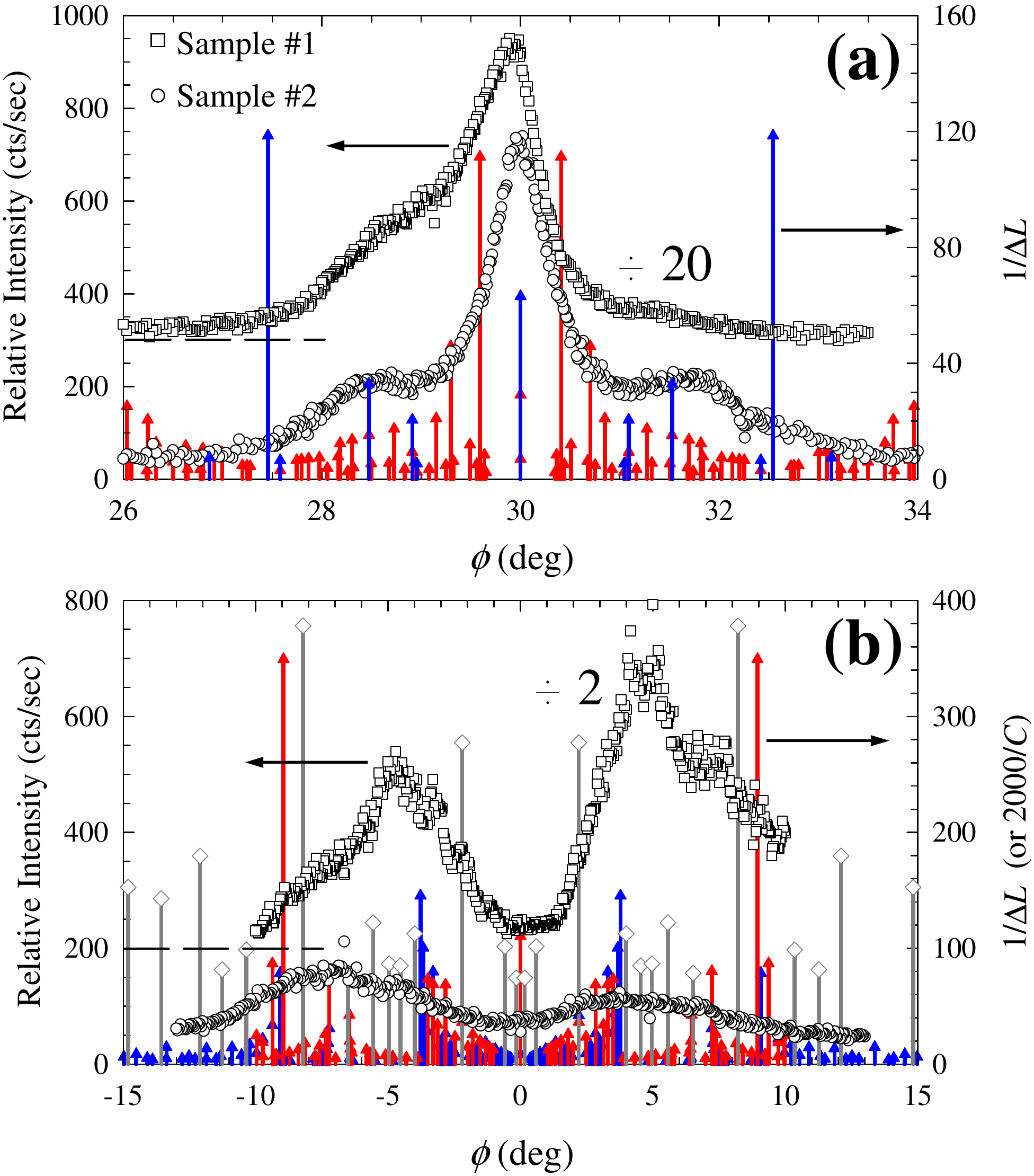}
\caption{X-ray angular distribution of graphene rotations near (a) $\phi=30$ and (b) $\phi=0$ for two different C-face samples. Sample $\#1$ ($\Box$) is a 30-layer film and ($\circ$) sample $\#2$ is a 6-layer film. Horizontal dashed line shows the offset for sample $\#1$ data. Drop lines show the position of graphene-SiC near commensurate structures for (red $\blacktriangle$) $L\!\times\! L$ and (blue $\blacktriangle$) $L\sqrt{3}\!\times\! L\sqrt{3}$ cells. The height of a drop line is $1/\Delta L$, where $\Delta L$ is the mismatch between the graphene structure and the SiC lattice. (Grey $\diamond$) show the position of rotated graphene sheets that are commensurate with a graphene sheet rotated $\phi=30^\circ$ relative to SiC (drop line height is proportional to $1/C$ for a $\sqrt{C}\!\times\!\sqrt{C}\text{R}\theta$ unit cell).}
\label{F:X-ray_ROT}
\end{figure}

Figure \ref{F:X-ray_ROT} shows the angular distribution of graphene planes near $\phi\!=\!0$ and $30^\circ$ for two different samples.  There are three things to point out in the figure.  First, as seen in the LEED, the angular distribution near $\phi\!=\!0$ is much broader than the distribution near $\phi\!=\!30$ (note the expanded $\phi$-scale in Fig.~\ref{F:X-ray_ROT}(b)).  The reason for this difference becomes obvious by considering the possible graphene structures that are nearly commensurate with SiC $(000\bar{1})$ when a rotated graphene plane is placed on the surface.\cite{Hass_JPhyCM_08} We have marked the angles for all $(L\!\times\! L)$ and $(L\sqrt{3}\!\times\!L\!\sqrt{3})\text{~R}30^\circ$ graphene-SiC near commensurate structures in Fig.~\ref{F:X-ray_ROT}.  For comparison, the height of the drop lines for the markers has been scaled by the inverse lattice mismatch between the graphene super cell and the SiC substrate: $1/\Delta L$.  This way of plotting accentuates structures with both small strain, $\epsilon$, and small cell sizes ($\Delta L=\epsilon L$). In addition to graphene being nearly commensurate with the SiC, two graphene sheets can be rotated relative to each other~\cite{Hass_PRL_08} to form a commensurate $(\sqrt{C}\!\times\!\sqrt{C})_\text{G}\text{R}\theta$ graphene-graphene super cell.\cite{Hass_JPhyCM_08} Figure \ref{F:X-ray_ROT}(b) marks the angular position of all graphene sheets that form a super cell with a graphene sheets rotated by $\phi\!=\!30^\circ$ with respect to the SiC.  The height of the drop lines for these graphene-graphene commensurate angles in Fig.~\ref{F:X-ray_ROT} has been drawn proportional to $1/C$ to highlight small unit cell structures.

It is clear that the density of graphene-SiC near commensurate rotations per unit of arc is higher for the $0^\circ$ rotations than for the $30^\circ$ rotations. In other words there are simply more possible commensurate structures oriented around $0^\circ$, with small energy differences.\cite{Kolmogorov_PRB_05,Campanera_PRB_07} The higher entropy associated with the distribution of commensurate angles near $\phi\!=\!0^\circ$ would explain why the SXRD angular distribution is broader in the $\phi\!=\!0$ azimuth.

The second thing to note in Fig.~\ref{F:X-ray_ROT} is that the exact distribution or rotation angles is sample-dependent. There are a number of possible reasons for this that can be related to the small energy differences for different rotational angles.  It is known on at least the Si-face that graphene grows out from substrate step edges.\cite{Hupalo_PRB_09} This means that a specific graphene orientation could be influenced by slight rotations as the graphene grows from the SiC step.  Since the substrate step direction depends on sample miscut and polishing, the distribution of graphene orientations would be sample dependent. The pleats in graphene that form when graphene is cooled from the growth temperature~\cite{Cambaza_Carbon_08} can be another source of small rotations that do not introduce defects in the hexagonal lattice of an otherwise continuous film.  Scanning tunneling microscopy (STM) has shown that rotational changes do occur at these topological boundaries.\cite{Biedermann_PRB_09}  Regardless of how the distribution is initially set, the rapid quench when the graphene is cooled from the growth temperature would freeze in the distribution.  Whether or not these rotations can be annealed out remains to be seen.

Finally, a detailed analysis of the SXRD angular distribution shows that the integrated area of the $\phi$ curves around $\phi\!=\!30^\circ$ is nearly the same as the area around $0^\circ$ (i.e $\int{I_{30}d\phi}/\int{I_0 d\phi}\sim 1\pm 0.2$).  In other words it is nearly as likely to find a graphene plane rotated $30^\circ$ relative to SiC as it is to have one rotated $0^\circ$.  This implies some order to the rotational stacking sequence and will be discussed in more detail in Sec.~\ref{S:Stacking}

Why the rotated graphene planes around $0^\circ$ form on the C-face graphene and are not produced during Si-face growth is not understood. It has been argued that the coupling between SiC and the first graphene layer is stronger on the Si-face compared to the C-face.\cite{Hiebel_PRB_08,Starke_JPCM_09} The existence of a stronger Si-face interaction is used to explain the observed difference in angular distribution on the two SiC surfaces.  The argument suggests that the stronger Si-face interaction forces the graphene to be aligned $\pm 30^\circ$ from the SiC, while the weaker C-face interaction allows the graphene to orient itself in multiple rotation directions.  Other groups, on the other hand, have argued for a stronger C-face interaction compared to the Si-face based on both graphene-SiC bond lengths and inverse photoemission data.\cite{Forbeaux_SS_99,Varchon_PRL_07,Mattausch_PRL_07,Hass_PRB_07,Hass_JPhyCM_08}

These arguments may be missing the point, because stronger or weaker interactions do not explain the different stacking order on the two surfaces.  In particular, a strong Si-face interaction does not explain why graphene grown on the Si-face is {\it AB} stacked.  Regardless of the interaction strength, the second layer must ``know'' that it should be rotated $60^\circ$ relative to the first.  It could be argued that the {\it AB} stacking pattern is formed by high temperature annealing of rotated sheets to the lower energy Bernal state. However, a similar ordering should, by the same logic, occur in C-face films that are grown at higher temperatures.  Since this does not happen, we are forced to propose something about the interface template that forces the {\it AB} stacking order in Si-face graphene. Similarly, a strong C-face interface interaction could explain why the graphene rotations angles in general match low strain and small unit cells near commensurate graphene-SiC structures.  However, it does not explain why C-face graphene grows with a more ordered rotational stacking sequence where planes are on average $30^\circ$ apart (as we'll see in Sec.~\ref{S:Stacking}). As in the Si-face case, a symmetry change must occur at the interface that cause the $30^\circ$ rotations to develop. This view makes the important question not which surface has the stronger interaction, but what is it about the different interface structures that imposes different rotational stacking sequences for graphene grown on the two SiC surfaces.

\subsection{Band Structure of C-face graphene\label{S:Band_C-face}}
\begin{figure}
\includegraphics[width=8.0cm,clip]{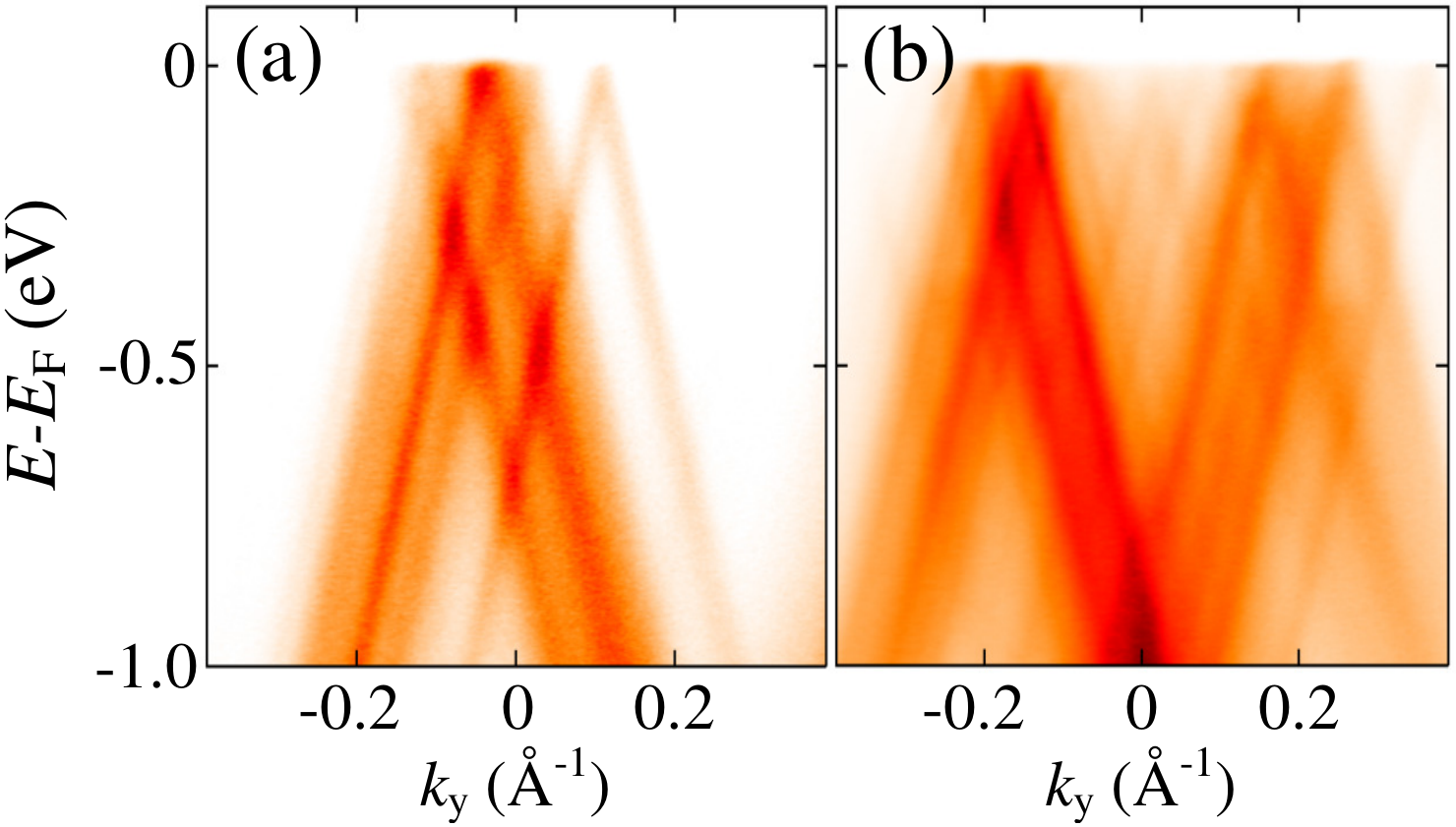}
\caption{ARPES scans taken at the $K$-point radius ($k_x=1.704\text{\AA}$) for a 10-layer grpahene film on the C-face of SiC. The photon energy is 36eV.  The scans are taken at two different emission directions: (a) along the SiC $\langle21\bar{3}0\rangle$ ($\alpha\!=\!0^\circ$) and (b) $\langle10\bar{1}0\rangle$ ($\alpha\!=\!30^\circ$) directions. The $k_y$ direction is defined in Fig.~\ref{F:Schem_LEED}.} \label{F:ARPES_planes}
\end{figure}

As discussed in Sec.~\ref{S:Symmetry}, the rotational stacking is expected to preserve the symmetry of an isolated graphene sheet.  Experimentally this means that the band structure of multilayer epitaxial C-face graphene should consist of a large number of multiply-rotated graphene Brillouin zones with the same distribution of rotation angles as seen in the SXRD data. Figure~\ref{F:ARPES_planes} shows ARPES scans at the $K$-point radius ($k_x\!=\!1.704\text{\AA}$) for electrons emitted along the SiC $\langle 21\bar{3}0\!\rangle$ and $\langle10\bar{1}0\rangle$ directions. Both scans show multiple linear dispersing Dirac cones but the distribution of these cones in the $\langle10\bar{1}0\rangle$ direction is bimodal and peaked at $\Delta k_y\!\sim\!\pm 0.2\text{\AA}^{-1}$ which, for small rotation angles, correspond to cones rotated $\alpha\!=\!30^\circ \pm\Delta\alpha\!\approx\!30^\circ\pm\tan^{-1}(\Delta k_y/k_{\Gamma K})\!=\!30^\circ\!\pm 6.7^\circ$.  Because the graphene $\Gamma K$ direction in ARPES is rotated $30^\circ$ from the graphene reciprocal space direction, $a^*_G$ [see Fig.~\ref{F:Schem_LEED}], the split distribution of graphene rotation angles observed in LEED around $\phi\!=\!0$ would produce a split distribution of $K$-points rotated around $\alpha\!=\!30$, i.e. the Dirac cone distribution would be centered around the $\langle10\bar{1}0\rangle$ direction and peaked at $\sim\!6-7^\circ$ as observed. Likewise the narrow distribution of graphene rotations around $\phi\!=\!-30$ would produce a narrow distribution of Dirac cones along the SiC $\langle 21\bar{3}0\!\rangle$ ($\alpha\!=\!0$) direction as seen in Fig.~\ref{F:ARPES_planes}.

\begin{figure}
\includegraphics[width=7cm,clip]{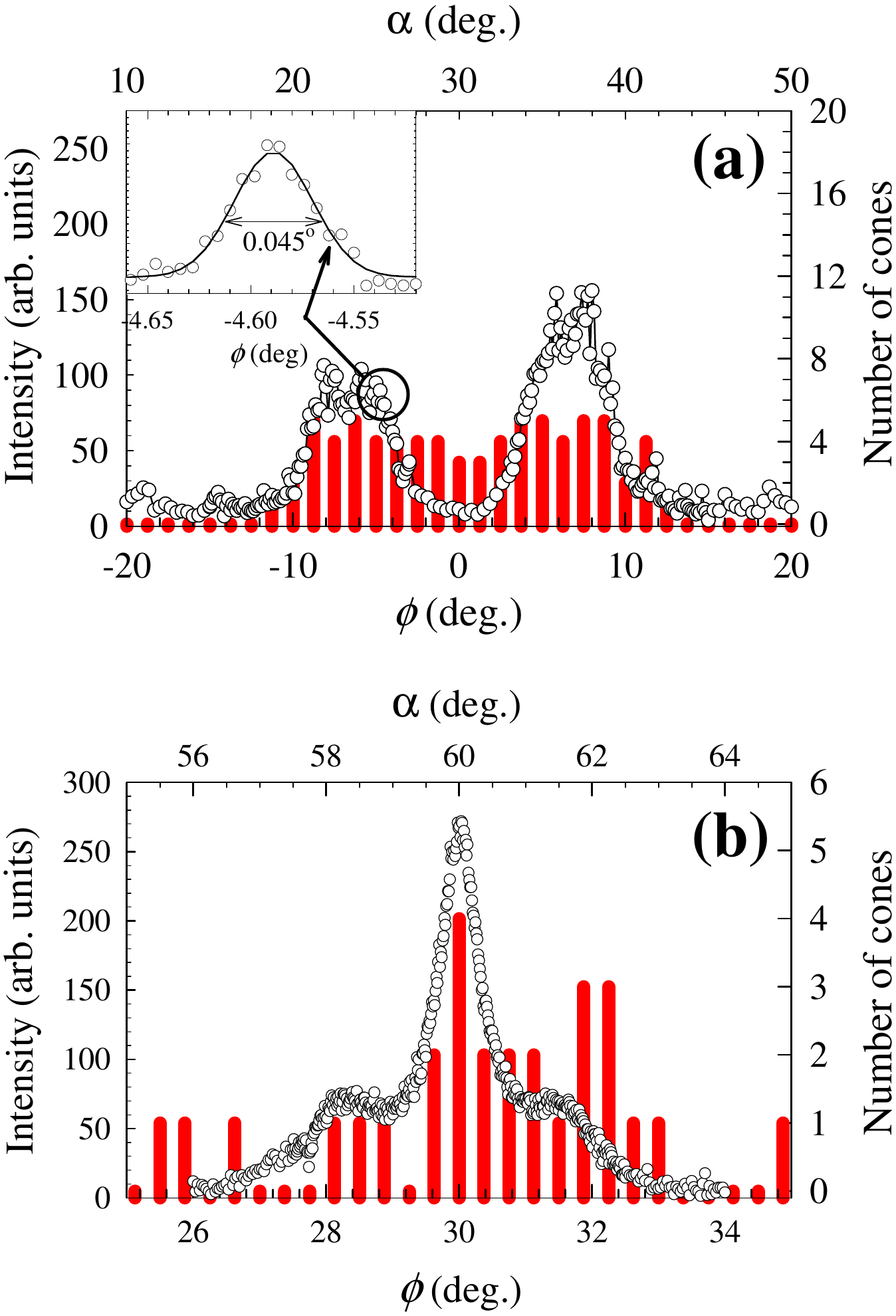}
\caption{ A comparison of the graphene SXRD angular distribution in $\phi$ ($\circ$) and the ARPES Dirac cone histograms in $\alpha$ (Solid lines). (a) SXRD graphene distribution taken around $\phi\!=\! 0$ and corresponding ARPES histogram of cones distributed around $\alpha\!=\!30$. (b) SXRD distribution taken around $\phi\!=\! 30$ and corresponding ARPES histogram of cones distributed around $\alpha\!=\!60$. Insert in (a) shows  a magnified view of a single rotation angle.}
\label{F:Diffraction}
\end{figure}

To emphasize the correlation between graphene rotation angle $\phi$ and the $\Gamma K$ rotation direction $\alpha$, we have marked the discrete rotation angles, $\alpha$, of the ARPES Dirac cones from the top three graphene layers against the bulk angular distribution of graphene rotations, $\phi$, measured by SXRD in Fig.~\ref{F:Diffraction}. It is clear that the ARPES cone positions correlate well with the data for graphene rotations in both the $\phi\!=\! 30^\circ$ and $0^\circ$ directions. Two experimental differences between SXRD and ARPES should be noted.  First the ARPES is only measuring cones in the upper 1-4 layers, while SXRD is measuring graphene planes throughout the entire film.  Second the SXRD beam size is 3mm while the ARPES beam size is $40\mu$m; this is why ARPES data shows a small number of discrete rotated cones and SXRD shows a more continuous distribution averaged over a large beam footprint. Also note that the angular width of each discrete rotation is very narrow; a detailed scan of one such angle is shown in the insert of Fig.~\ref{F:Diffraction}(a). Its width is $0.045^\circ$, corresponding to an x-ray rotational coherence distance of $\sim\!1\mu$m. This simply confirms that the x-rays can only measure graphene lengths up to the distance between SiC steps ($\sim\!1\mu$m for these samples) after which x-ray diffraction looses coherence as the graphene flows over the SiC substrate steps.\cite{Lauffer_PRB_08}

The discrete rotated set of Dirac cones clearly demonstrate that the rotated graphene planes have become electronically identical to isolated graphene sheets with nearly perfect linear dispersion at the $K$-point. Because the C-face films are thick, charge transfer from the SiC interface to the upper graphene layers is insignificant.  However, doping is observed in the graphene sheets at the graphene vacuum interface.  Doping is predominantly p-type but can vary to n-type depending on the sample.\cite{Sprinkle_PRL_09}  The reason for this sample-dependent doping is not known although weakly bound adsorbates could be the source. The doping ranges from $\sim\!33$meV p-doped on some samples to n-doped as low as -14meV on others. This gives a charge density that ranges between $\sim\!10^{11}-10^{10}\text{cm}^{-2}$, comparable to IR measurements from similar films ($5\!\times\!10^9\text{cm}^{-2}$).\cite{Orlita_PRL_08}

\section{Measurements of the Stacking Order\label{S:Stacking}}
So far we have shown that rotated graphene planes exist that have the band structure of an isolated graphene plane at the $K$-point. We have yet to discuss how these rotations planes are ordered. As we'll demonstrate, C-face graphene has an extremely high rotational ordering that makes it a new carbon allotrope.
\begin{figure}
\includegraphics[width=7.5cm,clip]{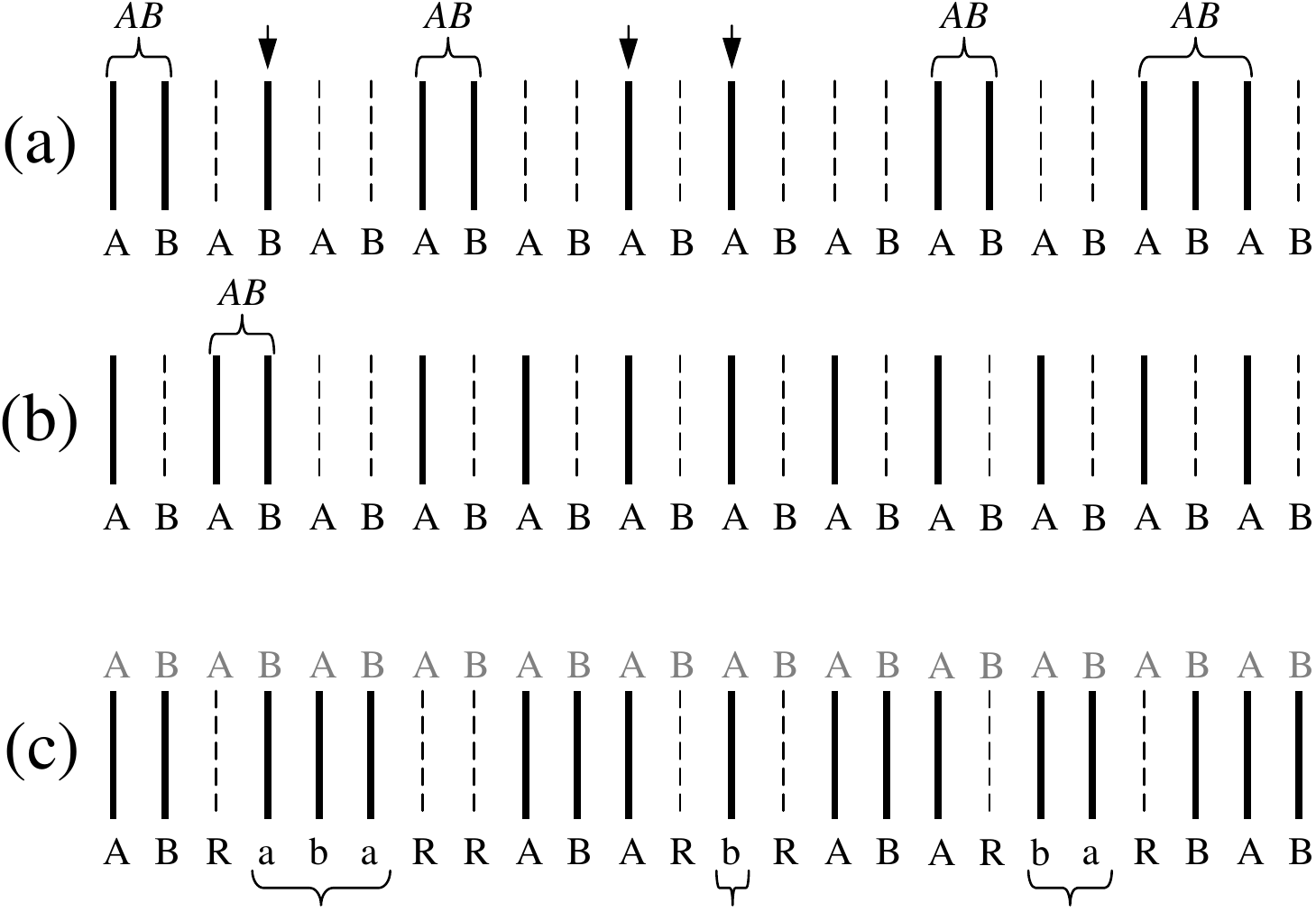}
\caption{Three rotational stacking models. (a) Bernal stacking sequence with 50\% randomly placed rotational planes (dashed lines). Arrows mark isolated {\it A} or {\it B} type graphene sheets. (b) Quasi-ordered stacking with the same concentration as (a) except the rotational plane sequence is nearly ordered. (c) Interrupted stacking sequence where the rotational planes cause a random phase boundary (marked by brackets) in the Bernal stacking sequence. Upper grey pattern in (c) represents the unperturbed sequence in (a) to show the difference between random stacking and the interrupted model.} \label{F:Stacking Fig}
\end{figure}

If one wishes to understand the transport in MEG films, it is the stacking order, and not just the concentration of rotational planes, that controls the fraction, $P_{AB}$, of planes in the film that are part of an {\it AB} stack. To see why this is so, we can compare two simple stacking models.  One model is to assume that rotational planes, $R$, are  randomly introduced into an otherwise {\it AB} stacked film with a probability $\gamma$. Figure \ref{F:Stacking Fig}(a) shows a random sequence of rotational planes.  Note that an $A$ plane must be adjacent to at least one $B$ plane to locally have the band structure of an {\it AB} pair [marked by brackets in Fig.~\ref{F:Stacking Fig}(a)]. $A$ or $B$ planes surrounded by rotational planes, [marked by arrows in Fig.~\ref{F:Stacking Fig}(a)], will still have the band structure of an isolated graphene sheet. For the random model the probability that any plane is part of an {\it AB} pair is determined by the two ways of retaining an {\it AB} pair.  This is either an ``..$ABR$..'' or ``..$ABA$..'' sequence.  $P_{AB}$ is then the sum of the probabilities that these two sequences exist:
\begin{equation}
P_{AB}= 2\gamma(1-\gamma)^2+(1-\gamma)^3=(1-\gamma)^2(1+\gamma)
\label{eq:AB_pair}
\end{equation}
In the random model there is always a chance that an $A$ or $B$ plane is surrounded by two rotated non-{\it AB} planes, even for very small $\gamma$'s so that $P_{AB}$ is always less than $1-\gamma$.

However, once the rotational plane stacking becomes ordered, Eq.~\ref{eq:AB_pair} is no longer valid and it will overestimate $P_{AB}$. In fact the overestimation can be very considerable. Consider the random model case with $\gamma=0.5$. Then on average there is a rotated plane every 2 graphene planes and $P_{AB}=0.375$.  However, if the distribution of rotational planes was completely ordered (i.e a rotation exactly every other plane as in Fig.~\ref{F:Stacking Fig}(b)), there would be \emph{no} {\it AB} pairs, since each $A$ or $B$ plane would be surrounded by a rotated plane.

In this section we discuss both x-ray diffraction and ARPES experiments that directly address the question of how the rotational planes in the MEG films are distributed. Early x-ray reflectivity measurements estimated the rotational plane probability to be approximately $\gamma\!\sim\!0.4$ based on an average expansion of the interplanar graphene lattice constant caused by $\pi^*$ bond interference in a random stacking model.\cite{Hass_PRB_07} This value of $\gamma$ would correspond to $P_{AB}\!\sim\!0.50$ in the random model. As we'll show in this section, this is a gross overestimate of the number of {\it AB} planes because the rotational stacking is more ordered than initially assumed. In fact the ordering is high enough that the {\it AB} planes, rather than the rotational planes, can be considered as faults in the stack.

\subsection{X-ray analysis\label{S:XRAY_anal}}
In order to demonstrate how SXRD data can be used to shed light on the stacking order in C-face graphene films, we begin by writing the general scattered x-ray amplitude from an N-layer graphene film as a function of the momentum transfer vector ${\bf q}$; ${\bf q}\!=\!{\bf k}_o\!-\!{\bf k}_i$ (where ${\bf k}_i$ and ${\bf k}_o$ are the incoming and scattered x-ray wave vectors, respectively),
\begin{equation}
A({\bf q}_\parallel , L)\propto e^{-(\sigma_\text{SiC}\pi L/c_G)^2}\sum^{N-1}_{k=0}F_k({\bf q}_\parallel)e^{i\pi Lc_k/c_G}.
\label{eq:Int_AB}
\end{equation}
The momentum transfer vector normal to the surface, $q_z$, has been written in terms of a variable $L$ defined as; $q_z=\pi L/c_G$ where $c_G=3.35$\AA~ is the graphite inter-layer spacing. $c_k$ is the vertical position of the $k^\text{th}$ graphene sheet. $F_k({\bf q}_\parallel)$ is the form factor for the $k^\text{th}$ graphene plane and depends on the rotation angle of the sheet. The gaussian term in Eq.~\ref{eq:Int_AB} accounts for the graphene RMS roughness, $\sigma_G$,  that results from the graphene draping over SiC substrate steps.\cite{Lauffer_PRB_08,Hass_PRB_07}

For the $(10L)$ crystal truncation rod (CTR),\cite{Robinson_CTR_86} the momentum transfer vector parallel to the surface is $|q_\parallel|\!=\!|a^*_G|\!=\!2\pi/a_G(\sqrt{3}/2)$. For the following discussions we assume that $\bf{a}^*_G$ is pointing along the SiC $\langle 10\bar{1}0\rangle$ direction so that $\phi\!=\!\pm 30$.  $F_k({\bf a}^*_G)$ can then take three possible values: it is equal to $F_A$ or $F_B$, the form factors of either an $A$ or $B$ graphene sheet corresponding to sheets rotated $\phi = 30^\circ$ and $-30^\circ$, respectively, or $F_k({\bf a}^*_G)\!=\!0$ corresponding to sheets rotated by $\phi\neq \pm 30^\circ$.

While a general analytic solution to Eq.~\ref{eq:Int_AB} is not possible for the $(10L)$ CTR, it can be solved for the case of Bernal stacking with random rotations like the model in Fig.~\ref{F:Stacking Fig}(a).  The probability of a non-Bernal rotation is defined as $\gamma$.  Note that there are three such types of possible rotations; an $A$ plane following another $A$ plane, a $B$ plane following another $B$ plane, and any plane that that is not an $A$ or $B$ plane.  The first two contribute intensity to the $(10L)$ CTR and the latter does not.  We define the probability that a non-Bernal plane is an $A$ or $B$ fault as $\beta$.  Using these definitions, the average intensity $I\!=\!AA^*$ from Eq.~\ref{eq:Int_AB} becomes;
\begin{subequations}
\begin{multline}
I({\bf q}_\parallel=a^*_G, L)\propto e^{-2(\sigma_\text{SiC}\pi L/c_G)^2}\left( B + \frac{\sin^2(\pi L N/2)}{\sin^2(\pi L)}\right. \\
\left.
\times\{ 2C-\left[ C-3(1-\gamma)\gamma\beta\right]\cos(\pi L)\right. \\
\left.
+\sqrt{3}\left[\gamma^2\beta^2-(1-\gamma)^2\right]\sin(\pi L) \}\right),\label{eq:Random_I}
\end{multline}
\begin{equation}
B(\gamma,\beta)=\gamma N[(1-\gamma)+\beta(1-2\gamma)],\label{eq:Random_back}
\end{equation}
\begin{equation}
C(\gamma,\beta)=[(1-\gamma)^2+\gamma^2\beta^2-(1-\gamma)\beta\gamma].\label{eq:Random_C}
\end{equation}
\label{eq:Random_All}
\end{subequations}
The intensity consists of a background term $B(\gamma,\beta)$ and a set of Bragg peaks at $L\!=$integer defined by the sinc function in Eq.~\ref{eq:Random_I}.

There are three features of this model that should be pointed out. First, there is a finite background that scales with $N$ and goes to zero when $\gamma\!=\!0$. Second, note that the full width at half maximum (FWHM), $\Delta L$, of the Bragg peaks is independent of $\gamma$ or $\alpha$ and only depends on the number of graphene layers; $\Delta L\!\sim\!2/N$.  The background and the $\gamma$ independent diffraction widths are common results for diffraction from random defects.\cite{Henzler_random} Finally, the random model predicts a specific relationship between the intensity of the $L\!=$ even and $L\!=$ odd Bragg points.  If we normalize the intensity in Eq.~\ref{eq:Random_I} by $\exp[-2(\sigma_\text{SiC}\pi L/c_G)^2]$, then the ratio of the normalized peak intensity above the background $I_p(L)=I(L)/\exp[-2(\sigma_\text{SiC}\pi L/c_G)^2]-B$ for $L$ odd and even is;
\begin{equation}
 \frac{I_p(L\!=\!\text{odd})}{I_p(L\!=\!\text{even})}=3\frac{(1-\gamma)^2+\gamma\beta[\gamma\beta-2(1-\gamma)]}
 {(1-\gamma)^2+\gamma\beta[\gamma\beta+2(1-\gamma)]}.
 \label{eq:1/0_Ratio}
\end{equation}
For random rotational planes in the {\it AB} stack, this ratio is 3 for any value of $\gamma$ as long as $\beta\!=\!0$. In other words, the random model requires {\it AA} or {\it BB} fault pairs to change the $I_p(L\!=\!\text{odd})/I_p(L\!=\!\text{even})$ ratio.

\begin{figure}
\includegraphics[width=7.5cm,clip]{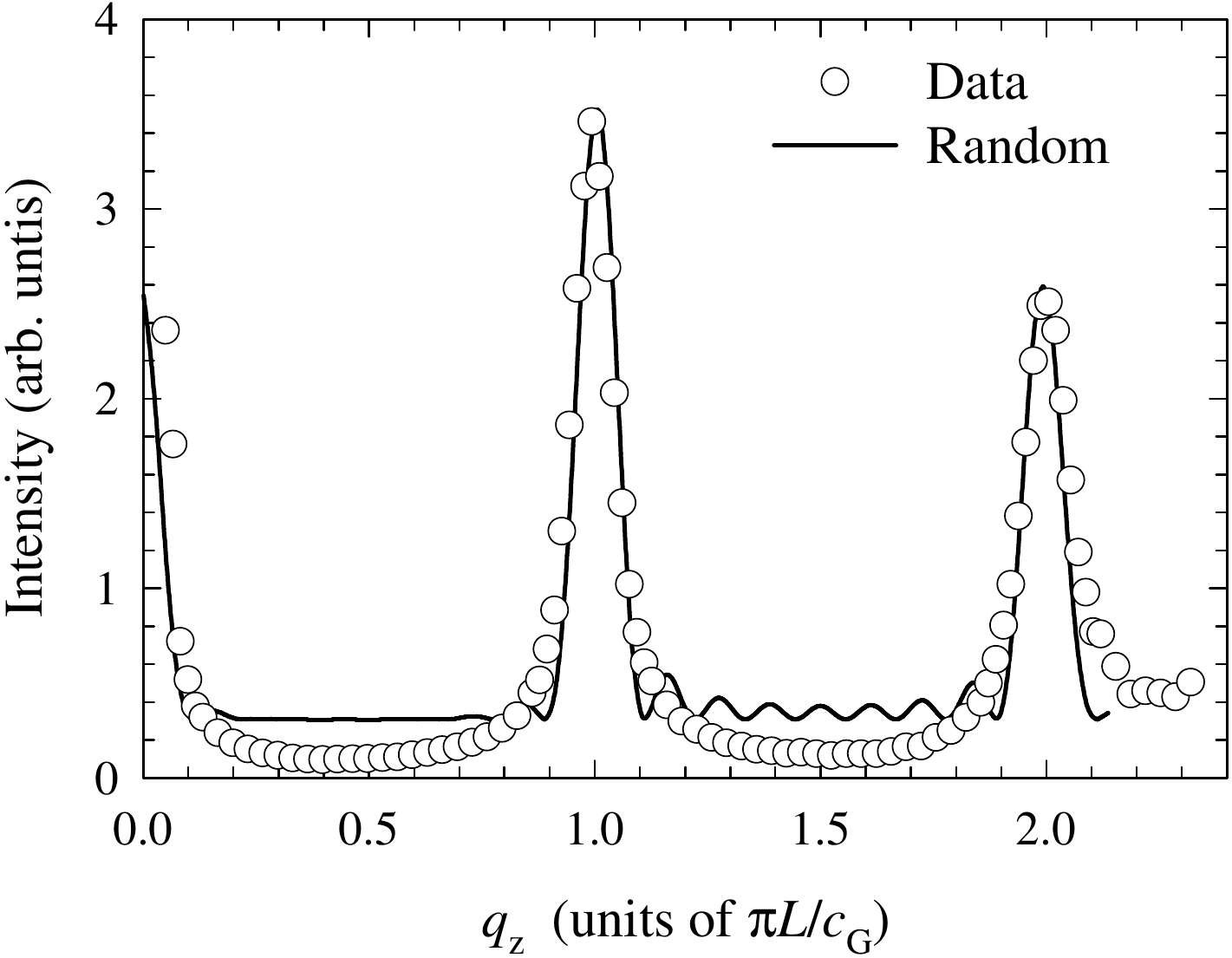}
\caption{The SXRD graphene $(10L)$ CTR from a 30-layer C-face graphene film.  The data ($\bullet$) has been normalized by the surface roughness term in Eq.~\ref{eq:Random_I} using $\sigma\!=\!0.54$\AA. The best fit random model X-ray reflectivity (line) using the $\gamma= 0.425$, $\beta = 0.26$, $N\!=\!18$ is shown.}\label{F:X-ray_Random}
\end{figure}

With these properties of the random model in mind, we can compare the experimental C-face graphene $(10L)$ CTR with Eq.~\ref{eq:Random_I}. Figure \ref{F:X-ray_Random} shows the experimental C-face graphene $(10L)$ intensity.  The data has been divided by $\exp[-2(\sigma_\text{SiC}\pi L/c_G)^2]$ using $\sigma_\text{SiC}\!=\!0.54\text{\AA}$ to correct for the SiC step roughness.  The obvious difference between the data and the fit is the significant background from the random model.  The magnitude of the background results from requiring the fit to match the experimental ratio of the odd and even Bragg peaks $I_p(\text{odd})/I_p(\text{even})\!=\!1.4$. We note that there are a range of $\gamma$ and $\beta$ that give similar fits to the ratio with the same high background. The fit in Fig.~\ref{F:X-ray_Random} was done using $\gamma\!=\!0.425$, and $\beta\!=\!0.26$. The value of $\gamma$ was chosen to be consistent with previous x-ray specular reflectivity estimates based on the interplanar graphene expansion for a random model.\cite{Hass_PRB_07} Regardless of the exact parameters in the fit, the small experimental background is the first indication that the rotational stacking is not random.

An even more significant problem with the random model is the number of graphene layers required to fit the data.  The best fit to the data in Fig.~\ref{F:X-ray_Random} that reproduces the Bragg peak widths is $N\!=\!18$.  However, ellipsometry measurements on the same film estimate $N$ to be nearly twice as large ($30\pm 3$).  Likewise, x-ray specular reflectivity ($q_\parallel\!=\!0$ in Eq.~\ref{eq:Int_AB}), which is sensitive to the total number of graphene planes regardless of the rotational distribution, measures a film thickness of $N\!=\!33\pm 2$.  This inability to fit the Bragg widths reflects the random models explicit lack of correlations in the rotational plane stacking.

In order to go beyond the random model and begin to understand the stacking in multilayer C-face films, requires the ability to compare the experimental data with other more complicated models where analytic solutions are not possible.  To do this we can generalize the CTR intensity calculations to a Quasi-ordered stacking (QOS) model. In this model the stacking sequence is generated by assuming two rotational planes are on average separated by $1/\gamma$ (the same as in the random model).  Disorder is introduced by allowing the rotational planes to occupy sites that are $\Delta n$ from the average according to gaussian weighted probability distribution; $\propto \exp[-(\Delta n-1/\gamma)^2/2w^2]$. The width of the gaussian ($w$) sets the degree of order: $w=0$ for perfect ordering and $w=\infty$ for random stacking. To be completely general the model also allows for another type of stacking disorder in addition to the rotational stacking.  We assume that the {\it AB} sequence can be disrupted after a rotational plane is introduced.  This is done by allowing a switch in the {\it AB} sequence from {\it ..ABRBA...} to {\it ..ABRab...} as illustrated in Fig.~\ref{F:Stacking Fig}(c). The switch occurs with a random probability, $\alpha$ ($\alpha\!=\!0$ means no switch occurs). Once a particular stacking sequence is set, Eq.~\ref{eq:Int_AB} is calculated and the CTR intensity is calculated.  The final calculated intensities are formed from an ensemble average of 1000 randomly generated distributions.
\begin{figure}
\includegraphics[width=7.5cm,clip]{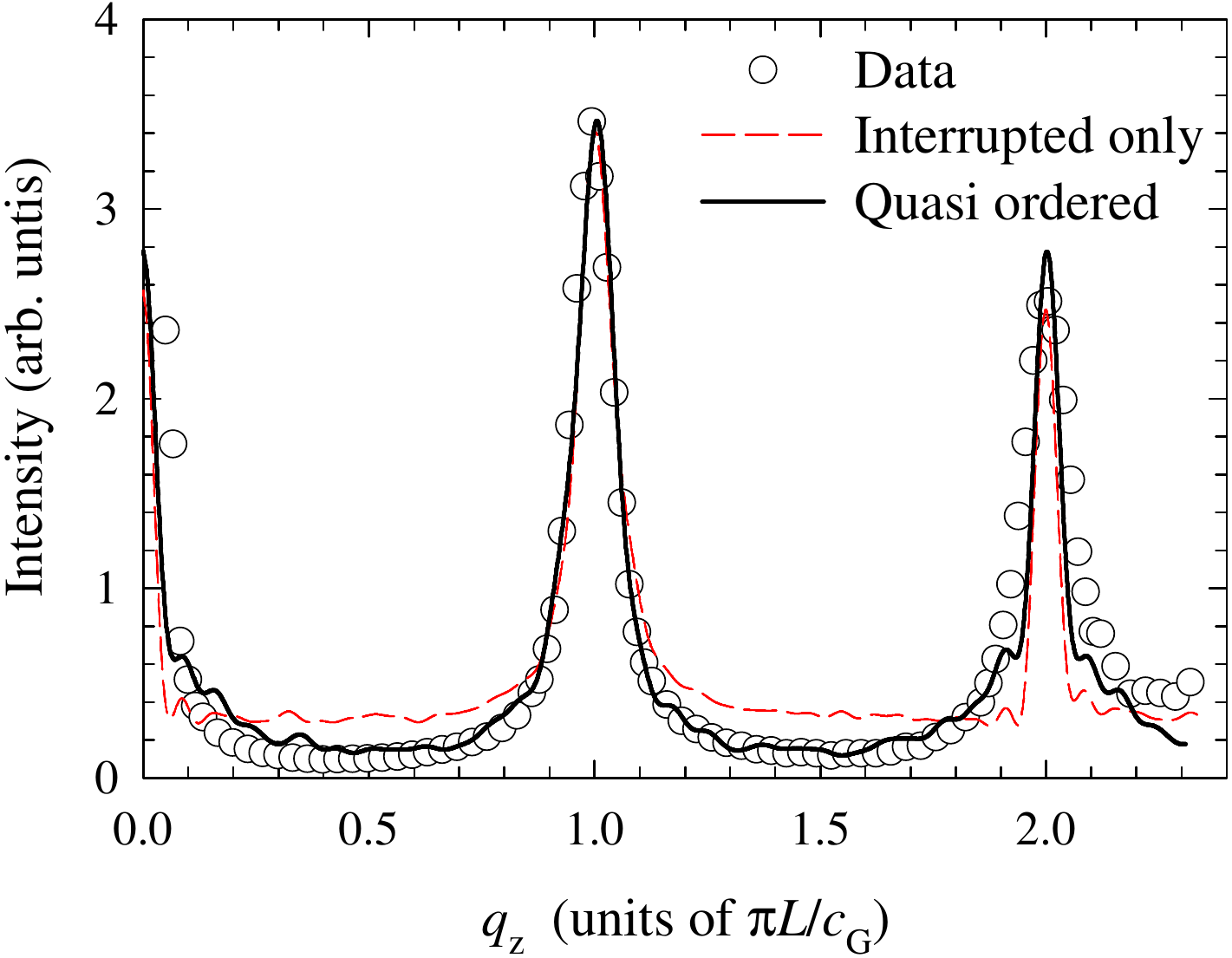}
\caption{Comparison of different model fits to the data. The dashed (red) line is a fit using random rotations ($w\!=\!0$) and interruption of the {\it AB} stacking ($\alpha\!>\!0$) as in Fig.~\ref{F:Stacking Fig}(c). This best fit uses $N\!=\!33$, $\gamma\!=\!0.545$ and a switching probability of $\alpha\!=\!0.1$. Solid (black line is the best fit using a quasi ordered model with switching. The fit uses $N\!=\!30$, $\gamma\!=\!0.55$, $w\!=\!0.05$ and a switching probability of $\alpha\!=\!0.09$.}\label{F:Stack_Calc_Comp}
\end{figure}

The interrupted stacking sequence is a logical extension of a purely random model.  A switch in the {\it AB} order will introduce phase boundaries in the stack that will broaden the Bragg peaks.  A fit to the graphene $(10L)$ data using the interrupted model is shown in Fig.~\ref{F:Stack_Calc_Comp}. The effect of switching does indeed broaden the $L$=odd Bragg peak width so that even a $N\!=\!33$ layer films (consistent with the value determined by SXRD specular reflectivity) can match the experimental width for a reasonable value of $\gamma$. Also note that unlike a purely random model, the ratio of odd and even Bragg peak intensities fit the experimental data without the need for any {\it AA} pairs (i.e. $\beta\!=\!0$). Beyond these points the random model, even with switching, still predicts a significant background not seen in the data. In addition the width of the $L$=odd Bragg peak widths in this model remains a function of $N$ only, just like the random model. It is clear that a more fundamental change to the stacking order is needed to fit the data.

Of the three models described in Fig.~\ref{F:Stacking Fig}, the quasi-ordered stacking model with a small amount of {\it AB} switching gives the best fit to the experimental data.  As shown in Fig.~\ref{F:Stack_Calc_Comp}, the model reproduces two features of the data that the random model cannot (even with {\it AB} switching); a low background and significant broadening of both the odd and even $L$ Bragg peaks.

In order to fit the low experimental background the order parameter was set to nearly zero $w\!=\!0.05$, with the average spacing between rotational planes set to $1/\gamma\!=\!1.83$.  In order to fit the $L\!=\! 1$ Bragg rod, $N$ was set equal to 30.  This is only slightly smaller than the thickness derived from the specular reflectivity data.  While the $L=$ even Bragg peaks have broadened significantly, the calculated rods are still narrower than the data.  This is because some degree of randomness is left in the model. The sharp gaussian distribution set by $w\!=\!0.05$ and the non-integer value of $1/\gamma$ requires that rotated planes landing at non-integer plane separations be randomly shifted to integer values (weighted to keep the rotational density equal to $\gamma$).

The most important results of this analysis is the low value of the order parameter.  It implies that the stacking sequence in multilayer epitaxial graphene is driven by predictive rather than stochastic processes. Presumably the SiC interface structure changes roughly every three SiC bilayers (the number of bilayers required to release the carbon necessary to form a single graphene layer).\cite{Hass_JPhyCM_08}  This structural change forces a rotation of a forming graphene sheet to be $30^\circ$ from the previously completed sheet.

\begin{figure}
\includegraphics[width=7.0cm,clip]{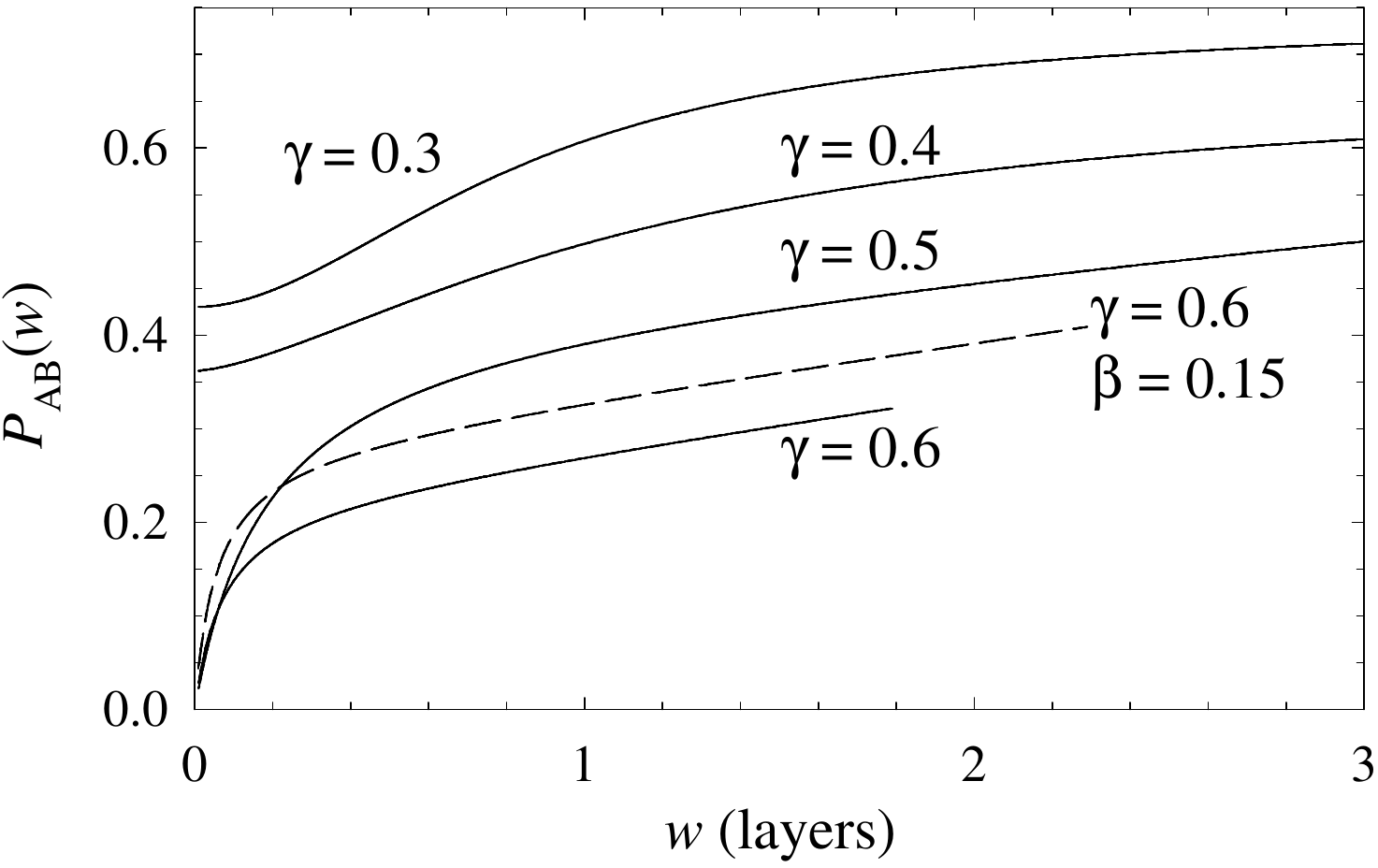}
\caption{A plot of $P_\text{AB}$ versus the order parameter $w$ for the quasi ordered model. The solid lines were calculated for a 10-layer film with $\alpha\!=\!0$ (no {\it AA} stacking) and $\beta\!=\!0$ (no {\it AB} switching).  The dashed line shows the effect of {\it AA} stacking on $P_\text{AB}$ for the case of $\gamma\!=\!0.6$.}\label{F:PAB_vs_Sigma}
\end{figure}

The low order parameter also predicts a low fraction of {\it AB} stacked planes.  Figure \ref{F:PAB_vs_Sigma} shows how the {\it AB} fraction depends on $w$.  For $\gamma$'s 0.5 and above, rotational stacking order causes $P_\text{AB}$ to fall to zero as $w\!\rightarrow\! 0$. The best fit to the graphene $(10L)$ rod data using the quasi ordered model with $\gamma\!=\!0.545$ leads to a value of $P_\text{AB}\!=\!19.8\%$. This low value of $P_\text{AB}$ shows that rather than thinking of the rotated planes as faults in an otherwise Bernal stacked film, MEG films should be viewed as rotationally ordered non-Bernal stacked films where the {\it AB} planes are themselves the faults. What is surprising is that, despite their small concentration, the {\it AB} planes in the film are placed as if they were in a Bernal stacked film, i.e. $\alpha$ is small.  This again suggests a periodic graphene formation mechanism at the SiC interface.

\subsection{ARPES analysis\label{S:ARPES_anal}}
ARPES data can also be used to corroborate the high degree of order inferred from the SXRD fits presented in the last section by estimating the relative number of {\it AB} planes in a C-face film. To understand how this is accomplished, we first review how the {\it AB} planes are identified and then discuss the experimental limitations of such measurements.

\begin{figure}
\includegraphics[width=7.5cm,clip]{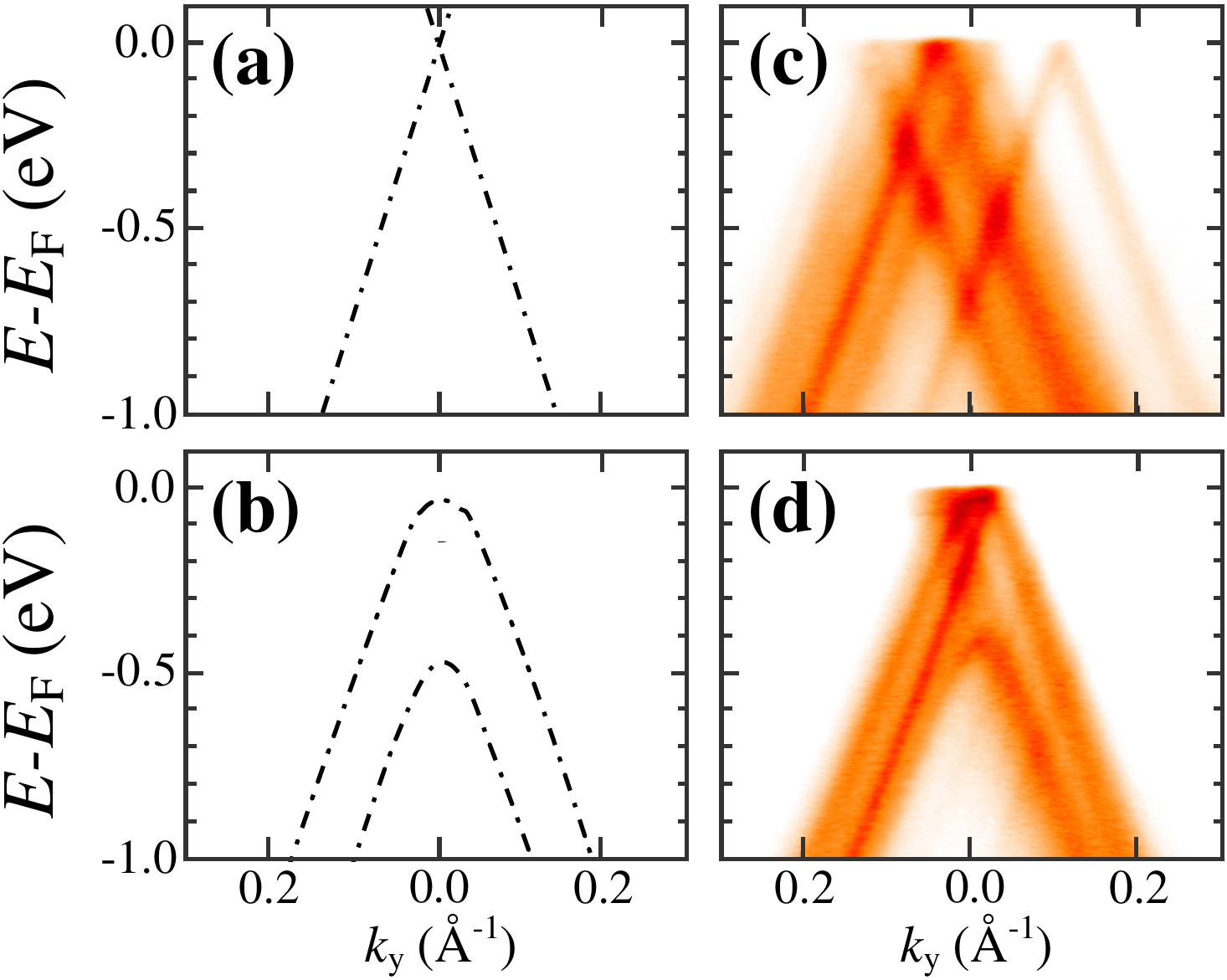}
\caption{Comparison of the ARPES band structure near the $K$-point for a single graphene sheet and an {\it AB} bilayer. The calculated tight binding dispersion from both (a) an isolated graphene sheet and (b) the band splitting from an {\it AB} stacked bilayer pair are shown.  (c) and (d) two experimentally measured bands for multilayered C-face graphene rotated $\phi\!=\!30^\circ$ from the SiC $\langle 21\bar{3}0\rangle$. (c) shows only linear graphene bands while (d) shows both linear bands and a split band associated with an {\it AB} stacked bilayer pair. The photon energy is 36eV and the photon beam size is $40\mu$m.} \label{F:ARPES_AB_Compare}
\end{figure}

The electronic signature of {\it AB} stacking is the splitting of the linear bands at the $K$-point.\cite{Orlita_PRL_08}  This is shown in Figs.~\ref{F:ARPES_AB_Compare}(a) and (b). When two graphene sheets are Bernal stacked, the single Dirac cone splits into two parabolic sub-bands.  The lower band is shifted nearly 0.5eV to lower binding energies. It is important to understand, as explained by Orlita et al.,\cite{Orlita_PRL_08} that the relative strength of the two bilayer bands is modulated by the perpendicular momentum transfer, $k_\perp$, defined as $k_\perp = (2m/\hbar^2) \sqrt{E_\text{kin}+V_0}$.  Here $E_\text{kin}$ is the measured electron kinetic energy and $V_0$ is the graphene inner potential, $V_0\!=\!16.5$eV.\cite{Zhou_AnPhys_06}  Ohta et al.,~\cite{Ohta_PRL_07} have shown that both sub-bands modulate with a period of $k_\perp\!\sim\!2.0\text{\AA}^{-1}$, which is not $2\pi/c_G\!=\! 1.86\text{\AA}^{-1}$ as expected. The relative strength of the upper sub-band (compared to the lower band) goes to nearly zero every $1\text{\AA}^{-1}$.

Figures~\ref{F:ARPES_AB_Compare}(c) and (d) shows experimental cuts perpendicular to the $\Gamma K$ direction (aligned with the SiC $\langle 10\bar{1}0\rangle$) to show the band structure from graphene planes rotated by $\phi\!=\!-30^\circ$.  Both directions show the typical series of linear dispersing Dirac cones. In addition to the linear Dirac cones, Fig.~\ref{F:ARPES_AB_Compare}(c) also shows the lower parabolic sub-band that corresponds to an {\it AB} stacked graphene bilayer. Because there are so many cones closely spaced in $k_y$ in the C-face films, distinguishing between linear and parabolic bands can be difficult. Because the lower sub-band is much lower in energy, it is much more easily distinguished from linear cones than the upper sub-band.  The problem of identifying {\it AB} bilayers can therefore be minimized by choosing a photon energy (and thus $k_\perp$) that makes the ARPES spectral weight nearly zero for the upper {\it AB} sub-band and at the same time maximizing the intensity of the lower sub-band. In principle then, the number of graphene planes that are part of {\it AB} pairs can be determined by counting the occurrence of these bi-layer bands relative to the number of linear bands from single graphene planes.

While a straight forward procedure, the limitations of this method must be kept in mind. First, for typical photon energies the photoelectron mean free path ranges between 1-5\AA).\cite{Barrett_PRB_05,Cumpson_SurfIA_97,Seah_SurfIA_79}  This means that ARPES only probes the stacking order in the first 3-4 layers.  Because the growth process of C-face films in the oven environment is not known, it is difficult to extrapolate surface values of $P_\text{AB}$ to the entire film.  Also, merely counting cones and disregarding the emission intensity can be problematic.  The intensity of a cone depends both on the depth of the graphene sheet in the stack and its relative area within the photon beam.  Counting cones alone allows the possibility of overweighing small relative area graphene sheets at the expense of large but deeper sheets. This effect can be minimized by using as small a probe area as possible.

With these provisos in mind, we have estimated $P_\text{AB}$ for a 9-10 layer graphene film using ARPES. To improve statistics, images similar to those in Fig.~\ref{F:ARPES_planes} were collected on up to six different locations on 2 different samples.  A total of 110 cones were measured.  Data was taken at a photon energy of 36eV to maximize the sensitivity to the lower bi-layer sub-band in Fig.~\ref{F:ARPES_AB_Compare}(b).  We find that the number of planes in {\it AB} pairs is $\sim\!15\%$ of the total planes in the film. Within a few percent, this fraction is the same for graphene planes rotated both $\phi\!=\!30^\circ$ and $0^\circ$ from the SiC $\langle 21\bar{3}0\rangle$ direction, indicating that there is no preference for Bernal stacking for film rotated in either principle direction. While an estimate, the ARPES value of $P_\text{AB}$ corroborate the SXRD results and places the concentration of {\it AB} stacked planes to be less than 20\% of the film.  Both techniques support the conclusion that the rotated graphene sheets are interleaved in the film and that the rotational stacking is ordered rather than random.

\section{Conclusion\label{S:Conc}}
Epitaxial graphene grown directly on both the SiC(0001) Si-face and $(000\bar{1})$ C-face has exceptional film quality.\cite{Hass_JPhyCM_08,Virojanadara_PRB_08,Emtsev_NatM_09,Tromp_PRL_09} These films are atomically flat with the graphene sheets being continuous over macroscopic distances (if not the entire crystal surface).  Furnace growth methods developed in the last 5 years are currently the best method to produce exceptionally large macroscopic graphene sheets. In fact STM studies have yet to find a single example of a discontinuous top layer in a C-face film. In this article, we show specifically how the unique structural and electronic properties of C-face epitaxial graphene continue to make this material a serious candidate for graphene electronics.

Unlike Si-face graphene where the stacking is {\it AB} or Bernal,\cite{Ohta_PRL_07} C-face graphene has an unusual stacking.\cite{Hass_PRL_08}  On this face the graphene sheets are not only oriented $30^\circ$ relative to the SiC $\langle 21\bar{3}0\rangle$ direction (typical of Si-face graphene), but are also rotated with a distribution of angles around $0^\circ$. The two orientations occur with equal likelihood. These rotated planes are interleaved in the film and are not part of isolated Bernal stacked films with multiple rotated domains like those in HOPG graphite. Furthermore, surface x-ray diffraction and angle resolved photoemission spectroscopy experiments show that the rotational plane stacking is highly ordered and not random. The exceptional large graphene sheet size and highly ordered rotational stacking of C-face epitaxial graphene demonstrate that this is a new form of graphene stacking and $\underline{\bf \text{not}}$ the disordered soot known as turbostratic graphite.\cite{Hutcheon_graphite}

The result of this type of ordered rotational stacking is that there are few adjacent planes in the film that are {\it AB} stacked.  The concentration of these Bernal stacked planes is less than 20\% meaning that they can be considered as faults in an otherwise ordered rotationally stacked film.  This unique stacking leads to a symmetry change in adjacent non-Bernal layers that makes these graphene sheets act electronically like a stack of isolated graphene sheets.\cite{LopesdoSantos_PRL_07,Latil_PRB_07,Hass_PRL_08}  Because these non-Bernal stacked layers make up the majority of the film, nearly the entire graphene film behaves electronically like a stack of isolated graphene sheets.\cite{Orlita_PRL_08,Miller_Science_09,Sprinkle_PRL_09}

While the reason for this type of rotational stacking is not understood in detail, the observed rotational distribution shows a strong correlation with graphene-SiC substrate commensurate structures that indicates an important orientational ordering mechanism at the SiC-graphene interface.

\begin{acknowledgments}
We wish to thank P.G. Baity for his help in preparing many of the figures in this article. This research was supported by the W.M. Keck Foundation, the Partner University Fund from the Embassy of France and the NSF under Grant No. DMR-0820382.  The Advanced Photon Source is supported by the DOE Office of BES, contract W-31-109-Eng-38.  The $\mu$-CAT beam line is supported by the US DOE through Ames Lab under Contract No.W-7405-Eng-82.
\end{acknowledgments}









\end{document}